\documentclass{llncs}

\usepackage{lmodern}
\usepackage[english]{babel}
\usepackage{booktabs}
\usepackage{url,xspace}
\usepackage{multirow}
\usepackage{verbatim}

\usepackage{tikz}
\usetikzlibrary{arrows.meta}
\usetikzlibrary{arrows}
\usetikzlibrary{backgrounds}
\usetikzlibrary{calc}
\usetikzlibrary{matrix}
\usetikzlibrary{positioning}
\usetikzlibrary{shapes}

\usepackage[T1]{fontenc}
\usepackage[utf8]{inputenc}
\usepackage{csquotes}
\usepackage{textcomp}
\usepackage{amssymb}
\usepackage{amsmath}

\usepackage{hyperref}

\hypersetup{
  colorlinks=false,
  hidelinks,
  pdfborder={0 0 0},
  bookmarksopen=true,
  bookmarksnumbered=true,
  breaklinks=true,
}

\usepackage{breakurl}

\usepackage{geometry}
\usepackage{bbm}

\makeatletter
\DeclareRobustCommand*{\bfseries}{%
  \not@math@alphabet\bfseries\mathbf
  \fontseries\bfdefault\selectfont
  \boldmath\let\mathbbnormal=\mathbb\let\mathbb=\mathbbm
}
\makeatother

\usepackage{pgfplots}
\pgfplotsset{compat=1.9}

\definecolor{plotcolor0}{RGB}{255,0,0}
\definecolor{plotcolor1}{RGB}{0,0,255}
\definecolor{plotcolor2}{RGB}{64,170,0}
\definecolor{plotcolor3}{RGB}{170,0,255}
\definecolor{plotcolor4}{RGB}{0,170,255}
\definecolor{plotcolor5}{RGB}{164,164,0}
\definecolor{plotcolor6}{RGB}{128,128,128}
\definecolor{plotcolor7}{RGB}{0,0,0}
\definecolor{plotcolor8}{RGB}{0,110,165}
\definecolor{plotcolor9}{RGB}{128,64,0}

\pgfplotscreateplotcyclelist{my_color}{%
  plotcolor0, every mark/.append style={solid,scale=0.8}, mark=x \\%
  plotcolor1, every mark/.append style={solid,scale=0.6}, mark=o \\%
  plotcolor2, every mark/.append style={solid,scale=0.8}, mark=diamond \\%
  plotcolor3, every mark/.append style={solid,scale=0.8}, mark=triangle \\%
  plotcolor4, every mark/.append style={solid,scale=0.8,rotate=180}, mark=triangle \\%
  plotcolor5, every mark/.append style={solid,scale=0.65}, mark=square \\%
  plotcolor6, every mark/.append style={solid,scale=0.8}, mark=star \\%
  plotcolor7, every mark/.append style={solid,scale=0.8}, mark=+ \\%
  plotcolor8, every mark/.append style={solid,scale=0.8}, mark=pentagon \\%
  plotcolor9, every mark/.append style={solid,scale=0.9}, mark=Mercedes star flipped \\%
  plotcolor0, every mark/.append style={solid,scale=0.8}, mark=x, dashed \\%
  plotcolor1, every mark/.append style={solid,scale=0.6}, mark=o, dashed \\%
  plotcolor2, every mark/.append style={solid,scale=0.8}, mark=diamond, dashed \\%
  plotcolor3, every mark/.append style={solid,scale=0.8}, mark=triangle, dashed \\%
  plotcolor4, every mark/.append style={solid,scale=0.8,rotate=180}, mark=triangle, dashed \\%
  plotcolor5, every mark/.append style={solid,scale=0.65}, mark=square, dashed \\%
  plotcolor6, every mark/.append style={solid,scale=0.8}, mark=star, dashed \\%
  plotcolor7, every mark/.append style={solid,scale=0.8}, mark=+, dashed \\%
  plotcolor8, every mark/.append style={solid,scale=0.8}, mark=pentagon, dashed \\%
  plotcolor9, every mark/.append style={solid,scale=0.9}, mark=Mercedes star flipped, dashed \\%
  plotcolor0, every mark/.append style={solid,scale=0.8}, mark=x, densely dotted \\%
  plotcolor1, every mark/.append style={solid,scale=0.6}, mark=o, densely dotted \\%
  plotcolor2, every mark/.append style={solid,scale=0.8}, mark=diamond, densely dotted \\%
  plotcolor3, every mark/.append style={solid,scale=0.8}, mark=triangle, densely dotted \\%
  plotcolor4, every mark/.append style={solid,scale=0.8,rotate=180}, mark=triangle, densely dotted \\%
  plotcolor5, every mark/.append style={solid,scale=0.65}, mark=square, densely dotted \\%
  plotcolor6, every mark/.append style={solid,scale=0.8}, mark=star, densely dotted \\%
  plotcolor7, every mark/.append style={solid,scale=0.8}, mark=+, densely dotted \\%
  plotcolor8, every mark/.append style={solid,scale=0.8}, mark=pentagon, densely dotted \\%
  plotcolor9, every mark/.append style={solid,scale=0.9}, mark=Mercedes star flipped, densely dotted \\%
}

\pgfplotsset{
  myPlot/.style={
    grid,
    width=59mm,height=48mm,
    major grid style={thin,dotted,color=black!50},
    minor grid style={thin,dotted,color=black!50},
    every axis/.append style={
      thin,
      tick style={
        line cap=round,
        thin,
      },
    },
    cycle list name={my_color},
    major tick length=3pt,
    minor tick length=1.5pt,
    legend cell align=left,
    xlabel near ticks,
    ylabel near ticks,
    title style={yshift=-0.6ex,font=\small},
  },
}

\pgfplotsset{
  log x ticks with fixed point/.style={
    xticklabel={
      \pgfkeys{/pgf/fpu=true}
      \pgfmathparse{exp(\tick)}%
      \pgfmathprintnumber[fixed relative, precision=3]{\pgfmathresult}
      \pgfkeys{/pgf/fpu=false}
    }
  },
  log y ticks with fixed point/.style={
    yticklabel={
      \pgfkeys{/pgf/fpu=true}
      \pgfmathparse{exp(\tick)}%
      \pgfmathprintnumber[fixed relative, precision=3]{\pgfmathresult}
      \pgfkeys{/pgf/fpu=false}
    }
  }
}

\newcommand{\arr}[1]{[#1]}
\newcommand{\set}[1]{\{#1\}}
\newcommand{\floor}[1]{\ensuremath{\lfloor #1 \rfloor}}

\usepackage{microtype}

\begin{document}

\title{
  COBS: a Compact Bit-Sliced Signature Index
}

\author{
  Timo Bingmann\inst{1}\and
  Phelim Bradley\inst{2}\and
  Florian Gauger\inst{1}\and
  Zamin Iqbal\inst{2}
}

\institute{
  Institute of Theoretical Informatics,\\
  Karlsruhe Institute of Technology, Germany
  \and
  European Molecular Biology Laboratory,\\
  European Bioinformatics Institute,\\
  Cambridge, United Kingdom
}
\maketitle


\begin{abstract}
  We present COBS, a COmpact Bit-sliced Signature index, which is a cross-over between an inverted index and Bloom filters.
  Our target application is to index $k$-mers of DNA samples or $q$-grams from text documents and process approximate pattern matching queries on the corpus with a user-chosen coverage threshold.
  Query results may contain a number of false positives which decreases exponentially with the query length.
  We compare COBS to seven other index software packages on 100\,000 microbial DNA samples.
  COBS' compact but simple data structure outperforms the other indexes in construction time and query performance with Mantis by Pandey et al. in second place.
  However, unlike Mantis and other previous work, COBS does not need the complete index in RAM and is thus designed to scale to larger document sets.
\end{abstract}

\section{Introduction}

In this paper we present an approximate $q$-gram index named COBS~\cite{gauger2018cobs}, short for COmpact Bit-sliced Signature index, which is a cross-over between an inverted index and Bloom filters.
The current focus of COBS is to index DNA and protein $k$-mers from sequencing experiments, but the data structure can also be used for indexing $q$-grams from other domains such as English text.

In living cells, DNA exists as long contiguous molecules, typically textually encoded as strings of \texttt{A}, \texttt{C}, \texttt{G}, and \texttt{T}.
Experimental methods for ``reading'' DNA have been developing rapidly; there are various approaches, but all involve breaking the DNA and ``reading'' (typically called ``sequencing'')  those fragments (these short strings are typically called ``reads'').
Read lengths started out moderately long (500--1000 characters) in the late 1990s, dropped down to 30 characters in 2008 with the advent of massively parallel technologies, and in the recent past, bleeding edge technologies have enabled reading of fragments as long as 1 million characters, albeit with a higher error rate.
\par
The output of sequencing experiments are stored both in raw format (text files of the read strings) and ``assembled format''  -- semi-heuristic best approximations to the underlying genome, also in text format, but of very variable quality, in particular when based on short read data.
Unambiguous reconstruction of the original string from the substrings is mathematically impossible unless the fragments are longer than the longest repeated substring. Another complication is that a great deal of data is generated by sequencing unknown mixtures of different genomes (e.g. mixtures of bacteria from within the human gut, or samples from humans infected by three different types of malaria parasite), making it very hard to reconstruct the underlying genomes.
\par
As sequencing technology has advanced, it has also become much cheaper and more widespread, and its output has been stored in publicly available archives, e.g. the European Nucleotide Archive (ENA) and the Sequence Read Archive (SRA) which maintain mirrors of all the data.
These archives now double in size every 18 months, and it is progressively more important to be able to search within the stored datasets, to find important genes or mutations, or combinations of mutations which are informative of function or ancestry.
All of these search queries can be expressed in terms of exact or approximate matching of strings.
In 2018, the ENA encompassed $1.5 \cdot 10^9$ microbial sequences and $8 \cdot 10^{15}$ base pairs  (i.e. characters) of read data~\cite{harrison2019european}, while the European Bioinformatics Institute reached 160\,PB of storage capacity~\cite{cook2019european}.


Despite the obvious similarities to standard document retrieval problems, the properties of DNA $k$-mer data are very different from traditional text corpora.
Google's index is reported to have in the order of $10^{13}$ documents containing $10^8$ unique terms~\cite{brin1998anatomy}, whereas the small benchmark set of 100\,000 microbial sequences used in our experiments already contain $2.2 \cdot 10^{10}$ distinct $31$-mers, of which $1.8 \cdot 10^{10}$ occur only once.
The frequency of terms in a natural language is power-law distributed, with underlying terms generated over hundreds of years, resulting in just a few new terms per document. Microbial genomes however encode many billions of years of evolution; each new genome generates thousands of novel $k$-mers.
There are also two other aspects whereby searching biological data differs from standard text retrieval. The first is that the index must support \emph{approximate queries} allowing detection of closely related DNA to the query.
Approximate pattern matching however is a notoriously difficult subject for text indices~\cite{navarro2001indexing,krugel2016approximate}.
The second is that users often want all hits, not just the top few as is typical in web search.

For COBS we chose the robust $q$-gram indexing approach~\cite{ukkonen1992approximate} and combined it with Bloom filters to reduce the term space size.
This can be considered a variant of \emph{signature files}, which have a long history in information retrieval~\cite{faloutsos1984signature} but were pushed to the sidelines for text search by inverted indexes~\cite{zobel1998inverted}.
Recently, they have been reconsidered as acceleration filters for large text search corpora~\cite{goodwin2017bitfunnel} by engineering them to adapt to the collection's characteristics.
With COBS we venture to combine signature files with one-sided errors introduced by Bloom filters and inverted files to design an ultra fast and scalable $q$-gram index which supports approximate queries delivering a small reasonable number of expected false positives.
Our contribution of making the signature files \emph{compact} first enables the index to be applied to corpora with highly varying document sizes, such as microbial DNA samples.

After reviewing related work in the following subsection, we present the new COBS index design in \autoref{sec:cobs-design}. In \autoref{sec:experiments} we then report on our experimental evaluation of COBS and seven other $k$-mer indexing software packages.

\subsection{Related Work}\label{sec:related-work}
Considering $q$-grams or $k$-mers of a sequence are a staple in bioinformatics~\cite{chikhi2019data}.

The earliest use of Bloom filters as an index for a collection of independent documents we could find is called \emph{Bloofi} by Crainiceanu and Lemire~\cite{crainiceanu2015bloofi}.
They propose to use a Bloom filter for each document and to arrange them either in a B-tree or as a \emph{Flat-Bloofi}.
The latter is similar to BIGSI and COBS without compaction.

The currently most cited line of work on DNA $k$-mer indices for approximate search are the \emph{Sequence Bloom Trees} (SBTs) first proposed by Solomon and Kingsford~\cite{solomon2016fast}.
In an SBT the $k$-mers of each document are indexed into individual Bloom filters, which are then arranged as the leaves of a binary tree.
The inner nodes of the binary tree are union Bloom filters of their descendants.
A query can then breadth-first traverse the tree, pruning search paths which no longer sufficiently cover a given threshold $\Theta$ of the query $k$-mers.

In the original SBT~\cite{solomon2016fast} a simple greedy clustering method is used, the bit union is stored in each inner node, and all nodes are RRR compressed~\cite{raman2002succinct} using SDSL~\cite{gog2014sdsllite}.
The first improvement, the \emph{Split Sequence Bloom Tree} (SSBT)~\cite{solomon2018improved}, splits the inner nodes into two Bloom filters: a \emph{similarity} filter and a \emph{remainder} filter, where the first contains all bits in both child filters and the second those set in either child minus the \emph{similarity} filter.
This representation allows descendant nodes to omit storing the bits in the \emph{similarity} filter explicitly, hence reducing space requirements while retaining the same information.

Simultaneously, Sun, Harris, Chikhi, and Medvedev proposed the \emph{AllSome Sequence Bloom Tree} (AllSome-SBT)~\cite{sun2018allsome}, which splits each inner node into an \emph{all} and a \emph{some} subfilter.
The \emph{all} filter contains bits in all leaves below the node, excluding those already set in the parent node, and the \emph{some} filter all bits in some leaves but not all.
Again, this representation allows exclusion of bits already known from the parent node's filters, and thus reducing space and enabling better compression.
Furthermore, the AllSome-SBT also improves on the clustering methods by employing an agglomerative hierarchical technique and by constructing batch Bloom filters for large query sets.

The currently smallest SBT variant is called \emph{HowDe Sequence Bloom Tree} (HowDe-SBT) by Harris and Medvedev \cite{harris2018improved}.
It decomposes the Bloom filters in each inner node into two bit vectors: the \emph{det} vector signals if a particular bit is \emph{determined} at this inner node, meaning that it is equal in all descendant leaves, and the \emph{how} vector signals if it is determined as zero or one.
All determined bits can be omitted from any children.
These two bit vectors are exactly the information needed to perform an efficient breadth-first search down the tree.
Furthermore, the authors introduce a \emph{culling} process to remove sparse inner nodes which don't reveal much information and thus speed up queries.

A completely different approach to indexing $k$-mers is taken by \emph{Mantis} from Pandey et al.~\cite{pandey2018mantis}.
In Mantis, a counting quotient filter (CQF)~\cite{pandey2017general} is used to construct a mapping from $k$-mers to \emph{color classes}, wherein $k$-mers with identical occurrence vectors for all documents are mapped to the same color class.
Incidence of color classes to documents can then be represented as a matrix, in which columns are associated with documents and each row corresponds to a color class.
Hence, bits set in the rows signal occurrence of any $k$-mer mapping to the color in the corresponding document list.
Mantis then compresses the bit vectors in the color matrix using RRR or with a spanning tree based approach.
The $k$-mer mapping is built from CQFs constructed by \emph{Squeakr}~\cite{pandey2018squeakr}, a $k$-mer counting tool.
Mantis differs from the other $k$-mer indexes referenced in this paper by being able to deliver \emph{exact} approximate matching results without false positives.

\emph{SeqOthello}~\cite{yu2018seqothello} is another $k$-mer index software package.
It contains an ``ensemble'' of encoding techniques for compressing the occurrence maps of $k$-mers in the document set.
Occurrence maps are then grouped depending on their density and encoding into disjoint buckets.
To locate the correct occurrence map for a $k$-mer, a hierarchy of \emph{Othello}s is built inside each bucket and over all bucket \emph{Othello}s.
An \emph{Othello}~\cite{yu2018memory} is a minimum perfect hash function mapping, which is fast and scalable but can introduce false positive results due to mapping of alien $k$-mers to random results.

\emph{BIGSI} (BItsliced Genomic Signature Index) by Bradley et al.~\cite{bradley2019ultrafast} is the direct ancestor of COBS and also a combination of Bloom filters and inverted indexes.
BIGSI however is a prototype programmed in Python and uses a key-value database such as BerkeleyDB or RocksDB as storage back-end.
It also does not contain the compaction feature introduced in COBS.

Related to $k$-mer indexing are \emph{colored de~Bruijn graph} representation data structures, which often contain an exact $k$-mer index but do not support approximate $k$-mer pattern searches.
The original implementation, \emph{Cortex} \cite{iqbal2012novo,iqbal2012high}, stored $k$-mers in a hash table, along with booleans for the four possible forward and backward edges in a single byte.
This was then followed by \emph{McCortex} \cite{turner2018integrating}, which added a second data structure to encode paths in the graph present in the original reads.
By contrast,  \emph{VARI}~\cite{muggli2017succinct}, \emph{Rainbowfish}~\cite{almodaresi2017rainbowfish}, and \emph{pufferfish}~\cite{almodaresi2018space} explore use of succinct data structures, the Burrows-Wheeler transform, and minimal perfect hash functions to save space and possibly even accelerate operations.
The \emph{Bloom Filter Trie} by Holley et al.~\cite{holley2016bloom} is another colored de~Bruijn graph representation based on the \emph{burst trie}~\cite{heinz2002burst}, wherein lookups for suffixes at compressed inner nodes are accelerated with Bloom filters.

\section{A Compact Bit-Sliced Signature Index}\label{sec:cobs-design}

In this section we present the index structure used in COBS.
We first generally review Bloom filters as a $q$-gram index in \autoref{sec:bf-of-signatures},
then turn to COBS' more compact bit-sliced representation in \autoref{sec:compact},
and discuss implementation details and algorithm engineering aspects in \autoref{sec:implementation}.

\subsection{Approximate Matching with Bloom Filters of Signatures}\label{sec:bf-of-signatures}

Given are an ordered set of \emph{documents} $\mathcal{D} = \arr{d_0,\ldots,d_{|\mathcal{D}|-1}}$, where each document $d$ is composed of a set of \emph{strings} $\set{t_{0},\ldots,t_{|d|-1}}$.
The number of items in a set or array is denoted with $|\cdot|$.
Each string $t$ is a zero-based array of $|t|$ characters from a finite ordered \emph{alphabet} $\Sigma$.
In the context of indexing DNA, the alphabet is usually $\{\texttt{A},\texttt{C},\texttt{G},\texttt{T}\}$, the documents are experiment samples, and the strings in each document can be reads or assembled genome sequences.
When indexing web sites, the alphabet may be the ASCII characters or English words, the documents could be web pages, and the substrings may be words, sentences, or paragraphs.

To facilitate approximate pattern matching we consider \emph{$q$-grams} of the strings \cite{ukkonen1992approximate}, commonly called \emph{$k$-mers} for DNA.
For each string $t$ with $|t| \geq q$ there are $|t| - q + 1$ consecutive substrings of length $q$.
For a document $d$, we denote with $G_q(d)$ the union of all $q$-grams in the strings in $d$.
Due to similarities with full-text indexing we also refer to the $q$-grams in a document as \emph{terms}.

A COBS index is composed of $|\mathcal{D}|$ Bloom filters~\cite{bloom1970space}, each representing an approximate membership data structure with one-sided error.  To construct a Bloom filter for a document $d$ we assume $k$ pairwise independent hash functions $h_0,\ldots,h_{k-1}$ with range $[0,w)$ and set the $k$ bits $h_i(s)$ in an array $f$ of $w$ bits for each $q$-gram $s \in G_q(d)$.
Testing for membership of a $q$-gram $s$ is performed by checking if all $k$ cells $h_i(s)$ are set, which can lead to false positives but never false negatives.

The entire document collection is thus represented by $|\mathcal{D}|$ bit arrays $\arr{f_0,\ldots,\linebreak f_{|\mathcal{D}|-1}}$, each a Bloom filter with possibly different parameters.
From previous work, the false positive rate $p$ of a Bloom filter of size $w$ with $k$ hash functions and $v$ inserted elements is known to be at most $(1 - (1-\frac{1}{w})^{kv})^k \leq (1 - e^{-kv/w})^k$.
Given a desired false positive rate $p$ and number of elements $v$, one can calculate a partial derivative of the last bound to determine good approximate parameters $k = \frac{w}{v} \ln 2$ and $w = - \frac{v \ln p}{(\ln 2)^2}$~\cite{broder2003network,mitzenmacher2005probability}.

To perform approximate matching for a pattern $P$, we follow previous work \cite{ukkonen1992approximate} and determine the $q$-gram distance of $P$ to all documents in the collection $\mathcal{D}$ by testing each of the query's $q$-grams $G_q(P)$ on all documents.
In COBS we present this positively as the $q$-gram \emph{score} of the query for each document.
The score is used to rank and return all documents containing at least a given percentage $K$ of the $|G_q(P)|$ terms in the query.

\begin{figure}[tb]
  \begin{minipage}[c]{0.48\textwidth}
    \newdimen\dummyDim%
    \begin{tikzpicture}[baseline]
      \begin{axis}[myPlot,
        xmin=0, xmax=10, ymin=0, ymax=1,
        domain=0:10, samples=100, no markers,
        xlabel={fill of Bloom filter $\frac{v}{w}$},
        ylabel={false positive rate $p$},
        xticklabel={\dummyDim=\tick pt
          \ifdim\dummyDim>0pt\pgfmathparse{1/\tick}\pgfmathprintnumber[fixed]{\pgfmathresult}\else1\fi},
        ]
        \addplot { (1 - exp(-1 * 1/x))^1 };
        \addplot { (1 - exp(-2 * 1/x))^2 };
        \addplot { (1 - exp(-3 * 1/x))^3 };

        \legend{$k=1$,$k=2$,$k=3$}
      \end{axis}
    \end{tikzpicture}
    \hspace{-3ex}
    \vspace{-1ex}
    \caption{Theoretical false positive rate of Bloom filters given fill and number of hash functions.}\label{sec:bf-fpr-fill}
  \end{minipage}
  \hspace{-2ex}\hfill%
  \begin{minipage}[c]{0.48\textwidth}
    \centering

\definecolor{col-green}{HTML}{6BF178}
\definecolor{col-red}{HTML}{FF5964}
\definecolor{col-blue}{HTML}{35A7FF}
\definecolor{col-yellow}{HTML}{FFE74C}

\colorlet{col-green-light}{white!40!col-green}
\colorlet{col-red-light}{white!40!col-red}
\colorlet{col-blue-light}{white!60!col-blue}
\colorlet{col-grey-light}{black!60}
\colorlet{col-grey-slight}{black!20}
\colorlet{col-yellow-light}{white!40!col-yellow}

\begin{tikzpicture}[
  baseline,
  font=\ttfamily,
  yscale=0.62,
  xscale=0.95,
  ]

  \filldraw[fill=white, draw=gray] (0,0)  rectangle ++(3,6);

  \filldraw[col-green-light, draw=gray] (0,1.2)  rectangle ++(3,0.3);
  \filldraw[col-green-light, draw=black] (0,4)  rectangle ++(3,0.3);
  \filldraw[col-green-light, draw=gray] (0,4.8)  rectangle ++(3,0.3);

  \node[draw=none, anchor=east] (data1)  at (-2, 3.7) {ACGA};
  \node[draw=none, anchor=east] (is1)    at (-2, 3.0) {CGAA};
  \node[draw=none, anchor=east] (great1) at (-2, 2.3) {GAAT};

  \draw[-{Latex}, draw=col-grey-light] (great1.east) .. controls +(east:1cm) and +(west:1cm) .. (0,1.35);
  \draw[-{Latex}, draw=col-grey-light] (is1.east) .. controls +(east:1cm) and +(west:1cm) .. (0,4.95);
  \draw[-{Latex}] (data1.east) .. controls +(east:1cm) and +(west:1cm) .. (0,4.15);

\end{tikzpicture}
    \caption{Access pattern of the classical bit-sliced index.}
    \label{fig:access-pattern-classic}
  \end{minipage}
\end{figure}

As Solomon and Kingsford already noticed for SBTs, in the case of approximate pattern search on Bloom filters, we are not interested in the false positive rate of a \emph{single} Bloom filter lookup.
Instead we are concerned with the false positive rate of a \emph{query} $P$.
More precisely, given $\ell = |G_q(P)|$ $q$-grams with the probability that more than $K \ell$ terms are false positives in the same filter.
\begin{theorem}[False Positive Rate of a Query, Theorem 2 in \cite{solomon2016fast}]\\
  Let $P$ be a query pattern containing $\ell = |G_q(P)|$ distinct terms. If we consider the terms as being independent, the probability that more than $\floor{K \ell}$ false-positive terms occur in a filter $f$ with false positive rate $p$ is
  $1 - \sum_{i=0}^{\floor{K \ell}} \binom{\ell}{i} p^i (1-p)^{\ell-i}$\,.
\end{theorem}
This theorem is derived by considering lookups of terms as independent Bernoulli trials and summing over the probability of zero to $\floor{K \ell}$ false positives among the $\ell$ trials, which yields a binomial distribution.
Given $K \geq p$, Solomon and Kingsford also apply a Chernoff bound and show that the false positive probability for a query to be detected in a document is $\leq \exp(- \ell (K - p)^2 / (2 (1 - p)))$.

These repeated trials into the Bloom filter allow us to push the false positive rate $p$ up higher than commonly used.
\autoref{sec:bf-fpr-fill} shows the false positive rate $(1 - e^{-kv/w})^k$ of Bloom filters depending on its fill $\frac{v}{w}$ and the number of hash functions $k$.
Traditional uses of Bloom filters for approximate membership queries consider an error rate of 0.01 or less and multiple hash functions as desirable.
Due to the inverse exponential relationship of a query's false positive rate with its length, coupled with the fact that more hash functions cost more cache faults or I/Os, the minimum $k=1$ and a high false positive rate around 0.3 are desirable for our $q$-gram index application.

For example, if we consider a query of length 100 containing $\ell = 70$ distinct $31$-grams, a false positive rate of $p = 0.3$, and threshold $K = 0.5$, then Theorem 1 yields a false positive rate of about 0.000143.
Which means there will be about 143 false positive results in one million documents on average.

\subsection{Bit-Slicing and Compaction}\label{sec:compact}

\begin{figure}[tb]
  \centering
  \input{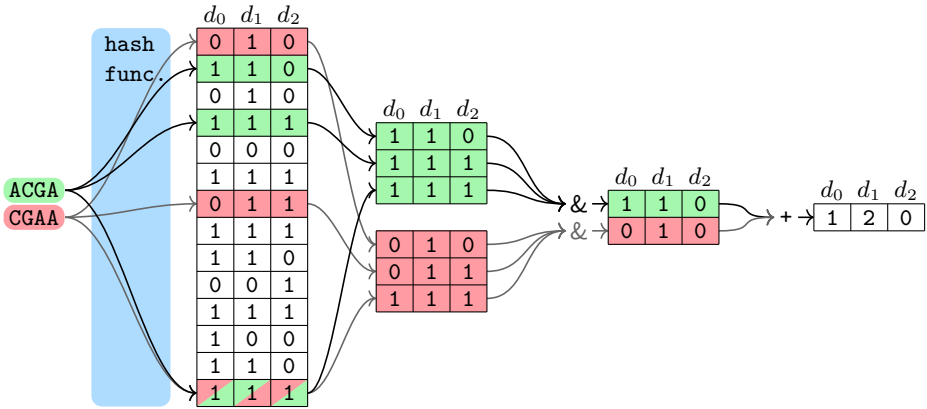}
  \vspace{-3ex}
  \caption{Architecture of the bit-sliced signature index and query processing steps.}\label{fig:cobs-signature-index}
\end{figure}

Provided all Bloom filters are of the same size $w$, one can store them as a $w \times |\mathcal{D}|$ bit matrix such that a row contains all bit cells at one index in the $|\mathcal{D}|$ filters (see left side of \autoref{fig:cobs-signature-index}).
This is also called a ``bit-sliced'' layout~\cite{wong1985bit} and was chosen for BIGSI and COBS to reduce the number of random accesses needed to evaluate a query.
Each row of a term can be scanned sequentially, as shown in \autoref{fig:access-pattern-classic}.
This is particularly important if the index is read from external memory, where scanning is much more efficient than random accesses.
The approach however requires all Bloom filters to use the same hash functions and be the same size.

\autoref{fig:cobs-signature-index} also illustrates how a query $P$ is performed using the bit-sliced Bloom filter matrix.
The $q$-grams of the query are hashed to determine the corresponding rows.
These $k |G_q(P)|$ rows are then scanned and an \emph{AND} join of $k$ rows is performed to determine which $q$-grams occur in which document. This yields an indicator bit vector ordered by document number. All indicator vectors are then added together to calculate the score for each document.
Only those documents reaching the query threshold $K |G_q(P)|$ are then reported as approximate matches.
Due to the one-sided error of the Bloom filters, only \emph{more} documents may be reported due to hash collisions; false negatives, i.e. missed hits, cannot occur.

One can also view the Bloom filter bit matrix as an inverted index: each row simply lists the document numbers containing the corresponding $q$-gram as indexes in a bit vector.
Unlike a traditional inverted index however, \emph{multiple} $q$-gram terms are superimposed in one row.
This leads to false positive matches.
In theory, one could apply all the methods developed by the information retrieval community~\cite{zobel2006inverted} to these bit vectors or posting lists.

The current version of a bit-sliced index however relies on all documents and resulting Bloom filters having the same size.
But larger documents result in denser bit vectors and smaller documents in sparser, as the number of bits set depends on the number of $q$-gram terms in the document.
Depending on the dataset, this creates vastly different false positive rates in the bit matrix.
Hence, we propose to \emph{adapt} the size of each Bloom filter bit array to the document it indexes and aim to keep the false positive rate \emph{constant}.
We call this a \emph{compact} bit-sliced signature index (the CO in COBS).

In theory one could adapt the Bloom filter size and hash function for each document.
In practice we want to store bits of rows as blocks of size $\Theta(B)$ in external memory, thus keep the parameters constant for $B$ consecutive documents.
Furthermore, instead of calculating a new hash function for each filter, we propose to use only one function with a larger output range and then use a modulo operation to map it down to each individual filter's size.
Both practical optimizations only incur a small deviation from the optimal index size and false positive rates.

\begin{figure}[tb]
  \begin{minipage}[c]{0.48\textwidth}
    \definecolor{stairs_plotcolor0}{RGB}{255,164,64}
    \definecolor{stairs_plotcolor1}{RGB}{0,0,255}
    \definecolor{stairs_plotcolor2}{RGB}{92,164,255}
    \pgfplotscreateplotcyclelist{stairs_color}{%
      stairs_plotcolor0, every mark/.append style={solid,scale=0.8}, mark=x \\%
      stairs_plotcolor1, every mark/.append style={solid,scale=0.6}, mark=o \\%
      stairs_plotcolor2, every mark/.append style={solid,scale=0.8}, mark=diamond \\%
    }
    \hspace{-3ex}%
    \begin{tikzpicture}[baseline]
      \begin{axis}[
        myPlot,
        enlargelimits=0.05,
        ymax=350,
        cycle list name={stairs_color},
        xlabel={documents\vphantom{$\frac{v}{w}$}},
        ylabel={Bloom filter size},
        x filter/.code={\pgfmathparse{#1/1000}\pgfmathresult},
        xticklabel={\pgfmathparse{\tick}\pgfmathprintnumber{\pgfmathresult}\,K},
        yticklabel={\pgfmathparse{\tick}\pgfmathprintnumber{\pgfmathresult}\,M},
        xtick={0,25,50,75,100},
        ]


        \addplot+[fill, no markers] coordinates { (0,0) (0,5031.34) (99999,5031.34) } \closedcycle;

        \addplot+[fill, no markers] coordinates { (0,0) (0,3.16386) (8191,3.16386) (8192,58.0916) (16383,58.0916) (16384,71.683) (24575,71.683) (24576,76.3739) (32767,76.3739) (32768,97.3663) (40959,97.3663) (40960,105.768) (49151,105.768) (49152,143.707) (57343,143.707) (57344,160.12) (65535,160.12) (65536,171.226) (73727,171.226) (73728,178.469) (81919,178.469) (81920,190.209) (90111,190.209) (90112,280.433) (98303,280.433) (98304,5031.34) (99999,5031.34) } \closedcycle;

        \addplot+[fill, no markers] coordinates { (99,0.0027448) (199,0.00952408) (299,0.0157566) (399,0.0211846) (499,0.0232256) (599,0.0268003) (699,0.0308797) (799,0.0370421) (899,0.0421448) (999,0.0480129) (1099,0.0548651) (1199,0.0610219) (1299,0.0674368) (1399,0.0752394) (1499,0.0840569) (1599,0.0914586) (1699,0.100097) (1799,0.109099) (1899,0.117917) (1999,0.126336) (2099,0.137091) (2199,0.148023) (2299,0.158957) (2399,0.169673) (2499,0.1801) (2599,0.189938) (2699,0.200583) (2799,0.214688) (2899,0.227846) (2999,0.239616) (3099,0.252521) (3199,0.266809) (3299,0.281166) (3399,0.293015) (3499,0.307229) (3599,0.321915) (3699,0.334817) (3799,0.350456) (3899,0.372176) (3999,0.391168) (4099,0.408591) (4199,0.430751) (4299,0.449157) (4399,0.468292) (4499,0.488666) (4599,0.508202) (4699,0.530654) (4799,0.558209) (4899,0.584597) (4999,0.617397) (5099,0.650564) (5199,0.686541) (5299,0.716975) (5399,0.747442) (5499,0.780977) (5599,0.816332) (5699,0.849022) (5799,0.894075) (5899,0.939416) (5999,0.984757) (6099,1.03563) (6199,1.09188) (6299,1.15446) (6399,1.23136) (6499,1.29799) (6599,1.36476) (6699,1.4444) (6799,1.5306) (6899,1.59868) (6999,1.69384) (7099,1.76951) (7199,1.85702) (7299,1.94799) (7399,2.05183) (7499,2.16592) (7599,2.27359) (7699,2.39094) (7799,2.53648) (7899,2.69025) (7999,2.84684) (8099,3.02637) (8199,3.178) (8299,3.34847) (8399,3.5389) (8499,3.7203) (8599,3.92671) (8699,4.12739) (8799,4.34491) (8899,4.58215) (8999,4.83345) (9099,5.09701) (9199,5.38947) (9299,5.74465) (9399,6.04298) (9499,6.40949) (9599,6.73603) (9699,7.06479) (9799,7.44071) (9899,7.80183) (9999,8.2376) (10099,8.74087) (10199,9.24834) (10299,9.76334) (10399,10.2884) (10499,10.8251) (10599,11.3844) (10699,11.9849) (10799,12.6355) (10899,13.2608) (10999,13.8362) (11099,14.4014) (11199,15.0625) (11299,15.6502) (11399,16.2962) (11499,16.981) (11599,17.7836) (11699,18.52) (11799,19.2045) (11899,20.0162) (11999,20.815) (12099,21.6202) (12199,22.5243) (12299,23.4141) (12399,24.2697) (12499,25.1333) (12599,26.1677) (12699,27.0588) (12799,28.0116) (12899,28.8468) (12999,29.6866) (13099,30.6689) (13199,31.5117) (13299,32.3616) (13399,33.3957) (13499,34.3336) (13599,35.2508) (13699,36.2088) (13799,37.3287) (13899,38.2941) (13999,39.2405) (14099,40.0993) (14199,40.9789) (14299,42.0675) (14399,42.7817) (14499,43.408) (14599,44.3694) (14699,45.5416) (14799,46.3634) (14899,47.4049) (14999,48.3646) (15099,49.2981) (15199,50.2315) (15299,50.9653) (15399,51.7752) (15499,52.6804) (15599,53.3399) (15699,54.1533) (15799,54.9932) (15899,55.7202) (15999,56.2743) (16099,56.9866) (16199,57.5026) (16299,57.8667) (16399,58.1185) (16499,58.315) (16599,58.4966) (16699,58.6952) (16799,58.8417) (16899,59.0013) (16999,59.1764) (17099,59.3569) (17199,59.5305) (17299,59.6908) (17399,59.8056) (17499,59.9172) (17599,60.0013) (17699,60.1202) (17799,60.2204) (17899,60.3047) (17999,60.4081) (18099,60.5028) (18199,60.5969) (18299,60.705) (18399,60.8196) (18499,60.954) (18599,61.0868) (18699,61.1988) (18799,61.3049) (18899,61.4179) (18999,61.5304) (19099,61.643) (19199,61.7575) (19299,61.8537) (19399,61.9798) (19499,62.0816) (19599,62.1802) (19699,62.2873) (19799,62.429) (19899,62.5808) (19999,62.7158) (20099,62.8456) (20199,62.9667) (20299,63.0956) (20399,63.2094) (20499,63.3084) (20599,63.4508) (20699,63.5878) (20799,63.731) (20899,63.8642) (20999,64.0539) (21099,64.2266) (21199,64.3959) (21299,64.6049) (21399,64.7957) (21499,64.9925) (21599,65.2005) (21699,65.3486) (21799,65.468) (21899,65.6054) (21999,65.7111) (22099,65.8302) (22199,65.9681) (22299,66.1228) (22399,66.3022) (22499,66.4609) (22599,66.6474) (22699,66.8419) (22799,67.0705) (22899,67.2895) (22999,67.458) (23099,67.5496) (23199,67.688) (23299,67.8662) (23399,68.1212) (23499,68.3447) (23599,68.6641) (23699,69.0264) (23799,69.4382) (23899,69.8062) (23999,70.269) (24099,70.7549) (24199,71.0526) (24299,71.3006) (24399,71.4552) (24499,71.5647) (24599,71.734) (24699,71.87) (24799,72.0192) (24899,72.1511) (24999,72.2959) (25099,72.4619) (25199,72.5881) (25299,72.7042) (25399,72.8474) (25499,72.9608) (25599,73.0797) (25699,73.1787) (25799,73.2564) (25899,73.317) (25999,73.3825) (26099,73.4422) (26199,73.5135) (26299,73.5783) (26399,73.6457) (26499,73.7008) (26599,73.7678) (26699,73.8354) (26799,73.8896) (26899,73.9415) (26999,73.9958) (27099,74.051) (27199,74.089) (27299,74.1224) (27399,74.1639) (27499,74.2003) (27599,74.2411) (27699,74.2872) (27799,74.3223) (27899,74.3707) (27999,74.4076) (28099,74.4505) (28199,74.4981) (28299,74.5423) (28399,74.581) (28499,74.6217) (28599,74.6606) (28699,74.6936) (28799,74.7318) (28899,74.7699) (28999,74.8145) (29099,74.8536) (29199,74.8929) (29299,74.9331) (29399,74.9851) (29499,75.0276) (29599,75.0697) (29699,75.1135) (29799,75.1499) (29899,75.1904) (29999,75.232) (30099,75.2754) (30199,75.3139) (30299,75.3488) (30399,75.3877) (30499,75.4289) (30599,75.4729) (30699,75.5117) (30799,75.5465) (30899,75.5918) (30999,75.6277) (31099,75.6705) (31199,75.7125) (31299,75.7502) (31399,75.7876) (31499,75.8218) (31599,75.8658) (31699,75.9054) (31799,75.9407) (31899,75.9864) (31999,76.0241) (32099,76.072) (32199,76.1197) (32299,76.1711) (32399,76.2159) (32499,76.2579) (32599,76.3035) (32699,76.3403) (32799,76.3819) (32899,76.4221) (32999,76.4633) (33099,76.5024) (33199,76.5536) (33299,76.5946) (33399,76.6369) (33499,76.6825) (33599,76.7231) (33699,76.7641) (33799,76.8079) (33899,76.8504) (33999,76.8931) (34099,76.9396) (34199,76.9796) (34299,77.02) (34399,77.0573) (34499,77.1012) (34599,77.1381) (34699,77.1803) (34799,77.2266) (34899,77.266) (34999,77.315) (35099,77.3503) (35199,77.3836) (35299,77.4298) (35399,77.4707) (35499,77.513) (35599,77.5576) (35699,77.6055) (35799,77.6646) (35899,77.7287) (35999,77.7881) (36099,77.8522) (36199,77.9202) (36299,77.9892) (36399,78.0632) (36499,78.1383) (36599,78.2319) (36699,78.3132) (36799,78.4027) (36899,78.4928) (36999,78.5885) (37099,78.696) (37199,78.7789) (37299,78.9167) (37399,79.0297) (37499,79.1647) (37599,79.3008) (37699,79.45) (37799,79.6163) (37899,79.8465) (37999,80.0641) (38099,80.3318) (38199,80.6616) (38299,81.0093) (38399,81.3396) (38499,81.7491) (38599,82.156) (38699,82.6104) (38799,83.0758) (38899,83.7591) (38999,84.3731) (39099,85.0206) (39199,85.6746) (39299,86.2814) (39399,87.1154) (39499,87.9704) (39599,88.6844) (39699,89.5617) (39799,90.2667) (39899,90.9057) (39999,91.4769) (40099,92.1774) (40199,92.9342) (40299,93.6005) (40399,94.2146) (40499,94.8528) (40599,95.6224) (40699,96.2937) (40799,96.8051) (40899,97.1577) (40999,97.4809) (41099,97.7854) (41199,98.0308) (41299,98.242) (41399,98.4649) (41499,98.6318) (41599,98.8186) (41699,98.9534) (41799,99.1101) (41899,99.2886) (41999,99.4238) (42099,99.5643) (42199,99.7128) (42299,99.8117) (42399,99.9259) (42499,100.033) (42599,100.129) (42699,100.246) (42799,100.345) (42899,100.458) (42999,100.558) (43099,100.662) (43199,100.758) (43299,100.852) (43399,100.95) (43499,101.01) (43599,101.091) (43699,101.182) (43799,101.262) (43899,101.342) (43999,101.421) (44099,101.485) (44199,101.542) (44299,101.613) (44399,101.707) (44499,101.795) (44599,101.867) (44699,101.936) (44799,102.022) (44899,102.105) (44999,102.211) (45099,102.295) (45199,102.38) (45299,102.468) (45399,102.542) (45499,102.591) (45599,102.633) (45699,102.711) (45799,102.793) (45899,102.886) (45999,102.968) (46099,103.045) (46199,103.132) (46299,103.206) (46399,103.307) (46499,103.4) (46599,103.507) (46699,103.603) (46799,103.702) (46899,103.793) (46999,103.869) (47099,103.971) (47199,104.081) (47299,104.174) (47399,104.247) (47499,104.312) (47599,104.381) (47699,104.452) (47799,104.517) (47899,104.6) (47999,104.702) (48099,104.791) (48199,104.872) (48299,104.958) (48399,105.039) (48499,105.132) (48599,105.227) (48699,105.296) (48799,105.406) (48899,105.525) (48999,105.628) (49099,105.724) (49199,105.821) (49299,105.909) (49399,106) (49499,106.13) (49599,106.231) (49699,106.348) (49799,106.428) (49899,106.537) (49999,106.666) (50099,106.795) (50199,106.935) (50299,107.041) (50399,107.138) (50499,107.235) (50599,107.363) (50699,107.567) (50799,107.749) (50899,107.955) (50999,108.135) (51099,108.343) (51199,108.498) (51299,108.692) (51399,108.873) (51499,109.065) (51599,109.22) (51699,109.351) (51799,109.525) (51899,109.704) (51999,109.895) (52099,110.038) (52199,110.196) (52299,110.377) (52399,110.439) (52499,110.477) (52599,110.596) (52699,110.819) (52799,111.022) (52899,111.258) (52999,111.498) (53099,111.789) (53199,112.053) (53299,112.397) (53399,112.832) (53499,113.208) (53599,113.526) (53699,113.955) (53799,114.535) (53899,115.178) (53999,115.95) (54099,116.801) (54199,117.791) (54299,118.725) (54399,119.205) (54499,119.769) (54599,120.444) (54699,121.56) (54799,122.441) (54899,123.114) (54999,124.235) (55099,125.286) (55199,126.109) (55299,127.011) (55399,127.914) (55499,128.773) (55599,129.663) (55699,130.518) (55799,131.583) (55899,132.665) (55999,133.968) (56099,134.952) (56199,135.958) (56299,137.252) (56399,138.157) (56499,138.965) (56599,139.811) (56699,140.558) (56799,141.061) (56899,141.577) (56999,142.134) (57099,142.629) (57199,143.023) (57299,143.476) (57399,143.947) (57499,144.33) (57599,144.83) (57699,145.183) (57799,145.512) (57899,145.819) (57999,146.023) (58099,146.2) (58199,146.423) (58299,146.554) (58399,146.685) (58499,146.918) (58599,147.237) (58699,147.604) (58799,147.968) (58899,148.357) (58999,148.735) (59099,148.904) (59199,149.16) (59299,149.477) (59399,149.977) (59499,150.393) (59599,150.811) (59699,151.282) (59799,151.725) (59899,152.076) (59999,152.503) (60099,152.922) (60199,153.243) (60299,153.609) (60399,153.941) (60499,154.243) (60599,154.5) (60699,154.753) (60799,154.994) (60899,155.216) (60999,155.44) (61099,155.622) (61199,155.789) (61299,155.949) (61399,156.098) (61499,156.283) (61599,156.449) (61699,156.617) (61799,156.73) (61899,156.868) (61999,156.991) (62099,157.072) (62199,157.163) (62299,157.267) (62399,157.347) (62499,157.443) (62599,157.547) (62699,157.664) (62799,157.812) (62899,157.959) (62999,158.083) (63099,158.192) (63199,158.269) (63299,158.348) (63399,158.418) (63499,158.485) (63599,158.549) (63699,158.622) (63799,158.693) (63899,158.76) (63999,158.83) (64099,158.892) (64199,158.967) (64299,159.035) (64399,159.11) (64499,159.191) (64599,159.264) (64699,159.339) (64799,159.41) (64899,159.474) (64999,159.549) (65099,159.636) (65199,159.727) (65299,159.84) (65399,159.951) (65499,160.069) (65599,160.218) (65699,160.347) (65799,160.489) (65899,160.675) (65999,160.874) (66099,161.084) (66199,161.338) (66299,161.6) (66399,161.858) (66499,162.093) (66599,162.308) (66699,162.559) (66799,162.836) (66899,163.099) (66999,163.432) (67099,163.74) (67199,164.001) (67299,164.284) (67399,164.506) (67499,164.744) (67599,164.955) (67699,165.121) (67799,165.25) (67899,165.371) (67999,165.522) (68099,165.648) (68199,165.772) (68299,165.908) (68399,166.029) (68499,166.113) (68599,166.201) (68699,166.294) (68799,166.402) (68899,166.541) (68999,166.663) (69099,166.805) (69199,166.967) (69299,167.109) (69399,167.258) (69499,167.438) (69599,167.628) (69699,167.764) (69799,167.922) (69899,168.084) (69999,168.242) (70099,168.393) (70199,168.52) (70299,168.641) (70399,168.754) (70499,168.851) (70599,168.955) (70699,169.082) (70799,169.181) (70899,169.285) (70999,169.384) (71099,169.482) (71199,169.596) (71299,169.691) (71399,169.787) (71499,169.882) (71599,169.974) (71699,170.082) (71799,170.182) (71899,170.261) (71999,170.346) (72099,170.43) (72199,170.507) (72299,170.575) (72399,170.647) (72499,170.716) (72599,170.775) (72699,170.824) (72799,170.866) (72899,170.908) (72999,170.95) (73099,170.986) (73199,171.023) (73299,171.059) (73399,171.093) (73499,171.139) (73599,171.175) (73699,171.214) (73799,171.255) (73899,171.301) (73999,171.356) (74099,171.412) (74199,171.46) (74299,171.509) (74399,171.556) (74499,171.598) (74599,171.652) (74699,171.706) (74799,171.758) (74899,171.82) (74999,171.875) (75099,171.937) (75199,172.003) (75299,172.056) (75399,172.111) (75499,172.173) (75599,172.23) (75699,172.29) (75799,172.344) (75899,172.412) (75999,172.464) (76099,172.525) (76199,172.592) (76299,172.674) (76399,172.749) (76499,172.817) (76599,172.898) (76699,172.975) (76799,173.038) (76899,173.115) (76999,173.18) (77099,173.25) (77199,173.35) (77299,173.434) (77399,173.529) (77499,173.631) (77599,173.722) (77699,173.811) (77799,173.898) (77899,174.012) (77999,174.087) (78099,174.178) (78199,174.289) (78299,174.405) (78399,174.504) (78499,174.607) (78599,174.715) (78699,174.804) (78799,174.915) (78899,175.001) (78999,175.092) (79099,175.198) (79199,175.303) (79299,175.43) (79399,175.541) (79499,175.662) (79599,175.78) (79699,175.905) (79799,176.024) (79899,176.144) (79999,176.256) (80099,176.385) (80199,176.511) (80299,176.631) (80399,176.754) (80499,176.917) (80599,177.031) (80699,177.145) (80799,177.252) (80899,177.354) (80999,177.463) (81099,177.589) (81199,177.696) (81299,177.804) (81399,177.923) (81499,178.035) (81599,178.133) (81699,178.243) (81799,178.367) (81899,178.452) (81999,178.545) (82099,178.63) (82199,178.719) (82299,178.798) (82399,178.9) (82499,179.013) (82599,179.104) (82699,179.181) (82799,179.251) (82899,179.321) (82999,179.387) (83099,179.477) (83199,179.561) (83299,179.656) (83399,179.766) (83499,179.845) (83599,179.944) (83699,180.04) (83799,180.133) (83899,180.217) (83999,180.345) (84099,180.441) (84199,180.567) (84299,180.643) (84399,180.74) (84499,180.828) (84599,180.914) (84699,181.014) (84799,181.127) (84899,181.252) (84999,181.365) (85099,181.501) (85199,181.648) (85299,181.782) (85399,181.922) (85499,182.067) (85599,182.212) (85699,182.345) (85799,182.483) (85899,182.635) (85999,182.769) (86099,182.929) (86199,183.073) (86299,183.199) (86399,183.363) (86499,183.528) (86599,183.683) (86699,183.847) (86799,184.015) (86899,184.178) (86999,184.318) (87099,184.482) (87199,184.697) (87299,184.894) (87399,185.096) (87499,185.257) (87599,185.412) (87699,185.599) (87799,185.77) (87899,185.958) (87999,186.156) (88099,186.354) (88199,186.494) (88299,186.669) (88399,186.839) (88499,187.01) (88599,187.257) (88699,187.427) (88799,187.612) (88899,187.763) (88999,187.908) (89099,188.075) (89199,188.282) (89299,188.516) (89399,188.675) (89499,188.865) (89599,189.052) (89699,189.27) (89799,189.491) (89899,189.723) (89999,189.923) (90099,190.184) (90199,190.409) (90299,190.651) (90399,190.903) (90499,191.126) (90599,191.337) (90699,191.596) (90799,191.84) (90899,192.078) (90999,192.284) (91099,192.518) (91199,192.742) (91299,192.956) (91399,193.176) (91499,193.374) (91599,193.558) (91699,193.734) (91799,193.938) (91899,194.114) (91999,194.291) (92099,194.543) (92199,194.784) (92299,195.003) (92399,195.257) (92499,195.528) (92599,195.806) (92699,196.083) (92799,196.415) (92899,196.728) (92999,197.093) (93099,197.443) (93199,197.8) (93299,198.24) (93399,198.573) (93499,198.905) (93599,199.189) (93699,199.541) (93799,199.927) (93899,200.48) (93999,200.941) (94099,201.412) (94199,201.978) (94299,202.571) (94399,203.131) (94499,203.712) (94599,204.303) (94699,204.911) (94799,205.655) (94899,206.259) (94999,207.013) (95099,207.726) (95199,208.632) (95299,209.586) (95399,210.657) (95499,211.778) (95599,213.574) (95699,215.327) (95799,217.695) (95899,219.905) (95999,222.567) (96099,224.839) (96199,226.532) (96299,227.855) (96399,229.989) (96499,231.767) (96599,233.659) (96699,235.668) (96799,237.335) (96899,239.62) (96999,241.572) (97099,244.286) (97199,246.649) (97299,248.798) (97399,250.695) (97499,253.431) (97599,255.553) (97699,257.287) (97799,258.582) (97899,260.293) (97999,262.466) (98099,267.293) (98199,273.04) (98299,279.903) (98399,289.09) (98499,301.232) (98599,310.585) (98699,325.551) (98799,338.749) (98899,355.155) (98999,386.958) (99099,420.887) (99199,492.559) (99299,566.29) (99399,690.752) (99499,846.785) (99599,1083.32) (99699,1111.59) (99799,1127.64) (99899,1616.3) (99999,5031.34) } \closedcycle;

      \end{axis}
    \end{tikzpicture}
    \vspace{-1ex}
    \caption{Compact index composed of small sub-indexes containing $B$ documents.}\label{fig:compact-steps}
  \end{minipage}
  \hfill%
  \begin{minipage}[c]{0.48\textwidth}
    \centering

\definecolor{col-green}{HTML}{6BF178}
\definecolor{col-red}{HTML}{FF5964}
\definecolor{col-blue}{HTML}{35A7FF}
\definecolor{col-yellow}{HTML}{FFE74C}

\colorlet{col-green-light}{white!40!col-green}
\colorlet{col-red-light}{white!40!col-red}
\colorlet{col-blue-light}{white!60!col-blue}
\colorlet{col-grey-light}{black!60}
\colorlet{col-grey-slight}{black!20}
\colorlet{col-yellow-light}{white!40!col-yellow}

\begin{tikzpicture}[
  baseline,
  font=\ttfamily,
  yscale=0.665,
  xscale=0.95,
  ]

  \filldraw[fill=white, draw=gray] (7,0) rectangle ++(1,1.5);
  \filldraw[fill=white, draw=gray] (8,0) rectangle ++(1,3.5);
  \filldraw[fill=white, draw=gray] (9,0) rectangle ++(1,6);

  \filldraw[col-green-light, draw=gray] (7,0.5) rectangle ++(1,0.3);
  \filldraw[col-green-light, draw=gray] (7,1.2) rectangle ++(1,0.3);
  \filldraw[col-green-light, draw=gray] (8,0.3) rectangle ++(1,0.3);
  \filldraw[col-green-light, draw=gray] (8,1.6) rectangle ++(1,0.3);
  \filldraw[col-green-light, draw=black] (8,2.5) rectangle ++(1,0.3);
  \filldraw[col-green-light, draw=gray] (9,1.2)  rectangle ++(1,0.3);
  \filldraw[col-green-light, draw=black] (9,4)  rectangle ++(1,0.3);
  \filldraw[col-green-light, draw=gray] (9,4.8)  rectangle ++(1,0.3);

  \filldraw[col-green-light, draw=black] (7,0.2) rectangle ++(1,0.3);

  \node[draw=none, anchor=east] (data2)  at (5, 3.7) {ACGA};
  \node[draw=none, anchor=east] (is2)    at (5, 3.0) {CGAA};
  \node[draw=none, anchor=east] (great2) at (5, 2.3) {GAAT};

  \draw[-{Latex}, draw=col-grey-light] (great2.east) .. controls +(east:1cm) and +(west:1cm) .. (7,0.65);
  \draw[-{Latex}, draw=col-grey-light] (is2.east) .. controls +(east:1cm) and +(west:1cm) .. (7,1.35);
  \draw[-{Latex}, draw=col-grey-light] (great2.east) .. controls +(east:2cm) and +(west:2cm) .. (8,0.45);
  \draw[-{Latex}, draw=col-grey-light] (is2.east) .. controls +(east:2cm) and +(west:2cm) .. (8,1.75);
  \draw[-{Latex}, draw=col-grey-light] (great2.east) .. controls +(east:2cm) and +(west:2cm) .. (9,1.35);
  \draw[-{Latex}, draw=col-grey-light] (is2.east) .. controls +(east:2cm) and +(west:2cm) .. (9,4.95);
  \draw[-{Latex}] (data2.east) .. controls +(east:1cm) and +(west:1cm) .. (7,0.35);
  \draw[-{Latex}] (data2.east) .. controls +(east:2cm) and +(west:2cm) .. (8,2.65);
  \draw[-{Latex}] (data2.east) .. controls +(east:2cm) and +(west:2cm) .. (9,4.15);

  \node (B) at (9.5,2.2) {$B$};
  \draw[->, shorten <=-3pt] (B) -- (9,2.2);
  \draw[->, shorten <=-3pt] (B) -- (10,2.2);

\end{tikzpicture}
    \vspace{-0.9ex}
    \caption{Access pattern of our compact bit-sliced index.}
    \label{fig:access-pattern-compact}
  \end{minipage}
\end{figure}

\autoref{fig:compact-steps} shows in light blue the desired Bloom filter size for the 100\,000 microbial documents used in our experiments ordered by size and with false positive rate 0.3 and one hash function.
The dark blue staircase function above the upward sloping curve shows batches of $B = 8\,192$ documents encoded with the maximum Bloom filter size of that block.
The visible dark blue area is the minor overhead for encoding documents block-wise.
If one uses only one Bloom filter size (the classic approach), then the index size would be the entire filled orange area, which extends upward to ensure the desired false positive rate for the largest document.

Due to the variance in size of microbial and other real-world documents, the \emph{compact} representation in COBS is essential.
In designing COBS, we also considered that today's SSD and NVMe storage technology now has orders of magnitude faster random access speeds~\cite{bingmann2019nvme} compared to rotational disks.
Thus with these new storage devices, the batched random access for many smaller blocks of size $B$, as used in the compact layout and illustrated in \autoref{fig:access-pattern-compact}, first becomes viable.

\subsection{Implementation and Engineering}\label{sec:implementation}

We implemented COBS as a command line search engine tool using C++ and plan to provide a Python interface to the underlying algorithm library.
The tool is open source and available from \url{https://panthema.net/cobs/}.
It can read DNA FASTA files, multi-document protein FASTA files, McCortex, or text files as documents and extract $q$-grams from them.
Depending on the format, the input data is broken into different $q$-gram sets: DNA reads are for example hashed independently, while English text is processed continuously.
We used xxHash~\cite{collet2014xxhash} for hashing the $q$-gram strings.
The $q$-grams or $k$-mers can optionally be canonicalized if their reverse complement are considered equivalent.

\textbf{Classic and Compact.}\quad
The COBS program can currently construct two index variants: \emph{classic} (ClaBS) and \emph{compact} (COBS).
In the classic index all documents are hashed using the same Bloom filter size, which depends on the desired false positive rate and the number of $q$-grams in the \emph{largest} document.
This is the non-compact version, which is similar to BIGSI, but was written for performance in C++ and with direct file accesses.
We will refer to it as \textbf{ClaBS} in the experimental results.

When constructing a compact index, the size of all documents are determined and the document set is reordered by size.
Then a subindex is constructed for every $B$ documents, as described in \autoref{sec:compact}.
Each subindex is actually a ClaBS index.
The subindices are simply concatenated into one large file.

While classic indexes with the same parameters can be concatenated straight-forwardly, compact indexes are more difficult to merge.
We may implement this in future versions of COBS by keeping some slack in the $\Theta(B)$ blocks and packing new documents into the best free block or by storing the subindices as separate files.
This would allow incremental augmentation of COBS compact indices.

\textbf{Parallelization.}\quad
Due to the massive amount of data to process, we parallelized construction and query for shared-memory systems.
ClaBS index construction we parallelized by building temporary indexes over batches of the documents and then merging them into larger indexes.
For compact index construction we parallelized construction of the subindices.

Pattern search in COBS can optionally be parallelized by processing disjoint partitions of the document scores in parallel and then selecting the top scores sequentially using a partial sort operation.

\textbf{Memory Mapped I/O.}\quad
For querying an index, we map the file into virtual address space using \texttt{mmap}.
The necessary rows of the inverted Bloom filter index are then read using simple memory transfers.
We experimented with directly issuing asynchronous I/O commands, but found only a negligible performance advantage that did not outweigh the higher code complexity.

Alternatively, COBS can also read the complete index into RAM and then run all queries.
This was added to compare performance against other indexing software which only work in RAM, e.g. Mantis, in \autoref{sec:experiments}.

\textbf{Single-Instruction Multiple-Data (SIMD).}\quad
Besides the I/O bottleneck, extracting the bits from the  index rows and adding them together required a considerable amount of running time in the query.

In the \emph{ADD} step of the query process (\autoref{fig:cobs-signature-index}), the rows are summed up to create the query result.
In this illustration we hid the fact that the rows that are output from the \emph{AND} step are bit-packed: each cell is represented by one bit.
In the output of the \emph{ADD} step, however, each document's score is represented by an integer specifying the number of matched query terms.
This poses a problem since the bits need to be unpacked before they can be processed.
Ideally we would like to unpack and process multiple bits at once.

We use a straight-forward mapping to expand 8 bits output by the \emph{AND} step to the $8
\cdot 16 = 128$ bits needed by the \emph{ADD} step when using 16-bit score counters.
This can be achieved by using one array lookup in a table of length 256 containing items of 128 bits.
With these 128 bits, the final result can then be calculated by summing up the expanded values for each document using a single 128-bit SIMD instruction.
The same approach can also be done with 32-bit score counters with 256-bit or two 128-bit instructions.

\section{Experimental Evaluation}\label{sec:experiments}

In this section we present a comprehensive evaluation of eight software packages for indexing $k$-mers from read or assembled genomic sequence data.

\begin{table}[t]
  \def\tabcolsep{3.5pt}
  \caption{Software, references, git hashes, and commit dates used in experiments.}\label{tab:software}
  \centering
  \begin{tabular}{lcl}
    Software / Index                 &                             & git hash and commit date                                     \\ \hline
    \textbf{SBT}                     & \cite{solomon2016fast}      & \texttt{977adfa} from March 1st 2019                         \\
    \textbf{SSBT} (Split-SBT)        & \cite{solomon2018improved}  & \texttt{710c95f} from July 10th 2018                         \\
    \textbf{AllSome-SBT}             & \cite{sun2018allsome}       & \texttt{4e1f2c5} from October 28th, 2018                     \\
    \textbf{HowDe-SBT}               & \cite{harris2018improved}   & \texttt{76e3c89} from March 1st, 2019                        \\
    \textbf{SeqOthello}              & \cite{yu2018seqothello}     & \texttt{68d47e0} from September 6th, 2018                    \\
    \textbf{Mantis}                  & \cite{pandey2018mantis}     & \texttt{3853c82} from January 29th, 2019                     \\
    \textbf{BIGSI}                   & \cite{bradley2019ultrafast} & \texttt{2ab35e5} from May 15th, 2019 using BerkeleyDB 4.8.30 \\
    \textbf{COBS} and \textbf{ClaBS} & [this]                      & \texttt{5328bd5} from May 24th, 2019                         \\
  \end{tabular}
\end{table}

\textbf{Software Packages.}\quad
We acquired copies of the original source code of seven other index software packages via Github.
The paper references, git hashes, and commit dates are listed in \autoref{tab:software}.
More information about each package can be found in the related work \autoref{sec:related-work}.
We compiled all software from source and additionally used ntCard~\cite{mohamadi2017ntcard} (v1.1.0) as a preprocessing step for the SBTs,
jellyfish~\cite{marccais2011fast} (v2.2.10) in other steps and as a library.

\textbf{Data.}\quad
Bradley et al.~\cite{bradley2019ultrafast} previously indexed the complete global corpus of microbial DNA data, some 450\,000 files. In doing this, they processed the raw data into $k$-mers.
Since this contains low frequency errors from the sequencing instruments, they ``de-noised'' it using standard methods from McCortex, and stored the remaining $k$-mers in
a binary format. We downloaded 100\,000 of these files from \url{http://ftp.ebi.ac.uk/pub/software/bigsi/nat_biotech_2018/ctx/}.
For microbial genomic read data $k$ was chosen as $31$, as this is large enough to (generally) guarantee uniqueness within a genome, without being so large as to frequently hit
a sequencing error. For scaling experiments we selected random subsets containing 100, 250, 500, 1\,000, 2\,500, 5\,000, 10\,000, 25\,0000, and 50\,000 documents from the 100\,000 base set, each contained in the larger subsets.
The 10\,000 document subset is the same as used in one of the BIGSI experiments~\cite{bradley2019ultrafast}.
The average document size is 42.77\,MiB stored in McCortex format, such that the entire 100\,000 microbial dataset is 3.984\,TiB in total.
Each document contains 3.4\,M 31-mers on average with the minimum being zero and the largest containing 138\,M 31-mers.
In total the 100\,000 dataset contains 336\,846\,M 31-mers to index.
While building the indexes using the various software all $k$-mers were included, without any occurrence threshold or cut-off.


For COBS' compact index $B = 1\,024$ documents were grouped into a subindex in the largest instance with 100\,000 documents.


\textbf{Platform.}\quad
We ran the experimental evaluation on a quad-socket Intel Xeon Gold 6138 2.0\,GHz 4$\,\times\,$20-core machine with 768\,GiB DDR4-2666 RAM and 4$\,\times\,$2\,TB NVMe Samsung 970 EVO SSD storage devices combined using RAID\,0.
The machine was running Ubuntu 18.04 with Linux kernel 4.15.0-48-generic and we used gcc 7.3.0.
The combined SSDs reached 12.2\,GiB/s sequential read, 2.3\,GiB/s sequential write, 741\,MiB/s random 4\,KiB block read, and 1\,188\,MiB/s random 4\,KiB block write speeds.

\textbf{Queries.}\quad
We designed four sets of batch queries to measure the performance of the indices, each set containing known true positives and true negatives in random order.
In each batch all queries are of the same length $\ell \in \{ 31, 100, 1\,000, 10\,000 \}$\,base pairs (bp).
To generate true positives, we first extracted all unitigs from the colored de~Bruijn graph representation of each document using McCortex, and then randomly chose queries from all $\ell$-grams in the unitigs.
To generate true negatives, we generated random query strings of length $\ell$, broke these down into $k$-mers, and checked that none of the $k$-mers were contained in any document.
To balance the size, we selected 100\,000 true positives and 100\,000 true negatives for $\ell = 31$ and $\ell = 100$, for $\ell = 1\,000$ we selected 10\,000 true positives and negatives each, and for $\ell = 10\,000$ we selected 1\,000 each.

The queries are stored in FASTA format and annotated with their origin (random negative or the correct document id).
After running the queries, we checked the results of each index software by comparing it against the true origin.
Using the true negatives in the $\ell = k = 31$ set we can determine the false positive rate of each index.

\textbf{Measurements.}\quad
To evaluate the software we measured many different performance metrics while running construction and the batch queries.
The machine was used exclusively when running the experiments.
Using interfaces from the Linux kernel, we measured wall-clock time, CPU user time which captures time spent computing in any user thread, the maximum resident set size (RSS) in memory as returned by the \texttt{time} utility, the number of bytes read and written to the SSDs in each step, and the change in storage usage.
We also recorded the resulting size of the index data files.

We flushed the disk cache before each build phase or query batch.
Each query batch was run three times: the first round started with a flushed (cold) cache, and the two subsequent rounds with a warm cache.
The rounds are labeled r0, r1, and r2.

\subsection{Results}




\begin{table}[t]
  \centering
  \caption{Construction wall-clock time, CPU time, memory usage, and resulting index size for 1\,000 microbial documents and all $k$-mer index software in our experiment}
  \label{tab:1000-documents}







  \gdef\sephline{
    \hline\hline
  }

  \def\tabcolsep{2.5pt}
  \begin{tabular}{r|*{9}{r}}
    & & & \hspace{-2ex}AllSome- & HowDe- & Seq- & & & & \\[-2pt]
    phase & SBT & SSBT & SBT & SBT & Othello  & Mantis & BIGSI & ClaBS & COBS \\
    \sephline
    & \multicolumn{9}{c}{Construction Wall-Clock Time in Seconds} \\ \cline{2-10}
       count & 2\,018 &  1\,974 & 1\,954 &  1\,959 &        &     &        &    &    \\
       bloom &    114 &     117 &    140 &     144 &    295 & 232 & 1\,881 &    &    \\
       build & 3\,097 & 21\,378 & 1\,401 & 68\,034 & 2\,225 & 987 & 2\,574 & 99 & 43 \\
    compress & 1\,768 &  5\,187 &     80 &  3\,802 &        &  45 &        &    &    \\
    \cline{2-10}
    total & 6\,996 & 28\,657 & 3\,576 & 73\,939 & 2\,520 & 1\,264 & 4\,455 & 99 & 43 \\
    \sephline
    & \multicolumn{9}{c}{Construction CPU (User) Time in Seconds} \\ \cline{2-10}
       count &  4\,574 &  4\,511 &  4\,475 &  4\,488 &         &         &          &        &        \\
       bloom & 11\,133 & 10\,967 & 10\,234 & 10\,278 & 28\,123 & 19\,162 & 169\,345 &        &        \\
       build &     855 &  5\,178 &     449 & 66\,872 &  2\,198 &     943 &   1\,767 & 1\,604 & 1\,430 \\
    compress &  1\,569 &  4\,832 &  1\,663 &  2\,857 &         &  3\,423 &          &        &        \\
    \cline{2-10}
    total & 18\,131 & 25\,489 & 16\,821 & 84\,495 & 30\,320 & 23\,527 & 171\,113 & 1\,604 & 1\,430 \\
    \sephline
    & \multicolumn{9}{c}{Construction Maximum RSS Memory Usage in MiB} \\ \cline{2-10}
       count &     518 &    518 &    518 &      518 &         &         &          &         &        \\
       bloom &     641 &    640 &    640 &      640 &     634 &  1\,756 &   4\,244 &         &        \\
       build & 11\,028 & 1\,523 & 7\,140 & 108\,147 & 12\,137 & 88\,357 & 246\,806 & 16\,245 & 2\,616 \\
    compress & 10\,953 &    992 &    560 &      963 &         & 16\,613 &          &         &        \\
    \cline{2-10}
    maximum & 11\,028 & 1\,523 & 7\,140 & 108\,147 & 12\,137 & 88\,357 & 246\,806 & 16\,245 & 2\,616 \\
    \sephline



    & \multicolumn{9}{c}{Index Size in MiB} \\ \cline{2-10}
    size & 19\,844 & 3\,254 & 21\,335 & 1\,911 & 4\,410 & 16\,486 & 27\,794 & 16\,236 & 3\,022 \\
    \hline

  \end{tabular}
\end{table}


In this section we present and discuss the results of our experiments with the eight index software packages.
The machine we selected for the experiments is a large server-class platform with 80 cores and large amounts of RAM.
While these properties are always good, we primarily chose it due to the 8\,TB of fast SSD storage, which is many times faster than traditional rotational disks.
For rapidly performing the experiments, this storage speed was crucial.

On the other hand the fast storage speed and massive multi-core processing power in our machine may highlight different aspects in the indexing software than previous comparisons.
Most prominently, algorithms which previously only had to process data rates known from rotational disks (100s of MiB/s) may become a bottleneck when dealing with SSD speeds (currently around 10~GiB/s).
Furthermore, most of the index software packages had no built-in provisioning for utilizing multi-core parallelism.
While we were able to accelerate embarrassingly parallel parts of the construction using bash (like creating Bloom filters for each file), in some software the main index build was still sequential.
On the other hand, one can argue that index construction time is not as important as query performance, but it still limits scalability.

\autoref{tab:1000-documents} shows our results from all eight software packages for only 1\,000 microbial DNA documents.
The steps in the construction of each index are shown as separate rows if it was possible to measure these independently.
We show both wall-clock time and CPU user time such that parallelized construction can highlight its speedup without obscuring the actual amount of computation.
\autoref{tab:1000-documents-queries} considers the time to run the query sets.
We only show wall-clock for queries due to space; all query computations are performed with a single thread such that this is a fair comparison.
Furthermore, for ClaBS and COBS the index is \emph{completely loaded} into RAM such that the comparison with the others is fair.
In future, it will become important to measure how many bytes were read from the disks per query, but in the current comparison we assume all index data is resident in RAM.


\begin{table}[t]
  \centering
  \caption{Query wall-clock time for 1\,000 microbial documents and all $k$-mer index software in our experiment}
  \label{tab:1000-documents-queries}
  \def\tabcolsep{2.5pt}
  \begin{tabular}{r|*{9}{r}}
    & & & \hspace{-2ex}AllSome- & HowDe- & Seq- & & & & \\[-2pt]
    phase & SBT & SSBT & SBT & SBT & Othello  & Mantis & BIGSI & ClaBS & COBS \\
    \sephline



    $\ell$ \phantom{bp r0} & \multicolumn{9}{c}{Query Wall-Clock Time in Seconds} \\ \cline{2-10}
       31 bp r0 &  31 &     80 &  20 &  34 & 62 & 12 & 281 & 10 &  8 \\
       31 bp r2 &  26 &     76 &  19 &  33 & 62 & 13 & 289 &  9 &  8 \\
      100 bp r0 & 663 & 3\,183 & 100 & 600 & 73 & 22 & 783 & 14 &  9 \\
      100 bp r2 & 649 & 3\,153 &  95 & 588 & 73 & 23 & 455 & 14 &  9 \\
     1000 bp r0 & 794 & 3\,466 & 112 & 670 & 63 & 21 & 660 & 15 & 10 \\
     1000 bp r2 & 781 & 3\,435 & 108 & 659 & 64 & 27 & 310 & 13 & 10 \\
    10000 bp r0 & 802 & 3\,273 & 112 & 622 & 62 & 23 & 699 & 16 & 11 \\
    10000 bp r2 & 790 & 3\,243 & 111 & 613 & 62 & 22 & 316 & 15 & 11 \\
    \cline{2-10}
    total r0--r2 & 6\,775 & 29\,833 & 1\,007 & 5\,710 & 783 & 252 & 5\,177 & 154 & 114 \\
    \sephline



    & \multicolumn{9}{c}{Document False Positive Rate for 31 bp Queries} \\ \cline{2-10}
    rate & 0.004 & 0.004 & 0.004 & 0.004 & 0.001 & 0.000 & 0.027 & 0.024 & 0.227 \\
    \hline
  \end{tabular}
\end{table}

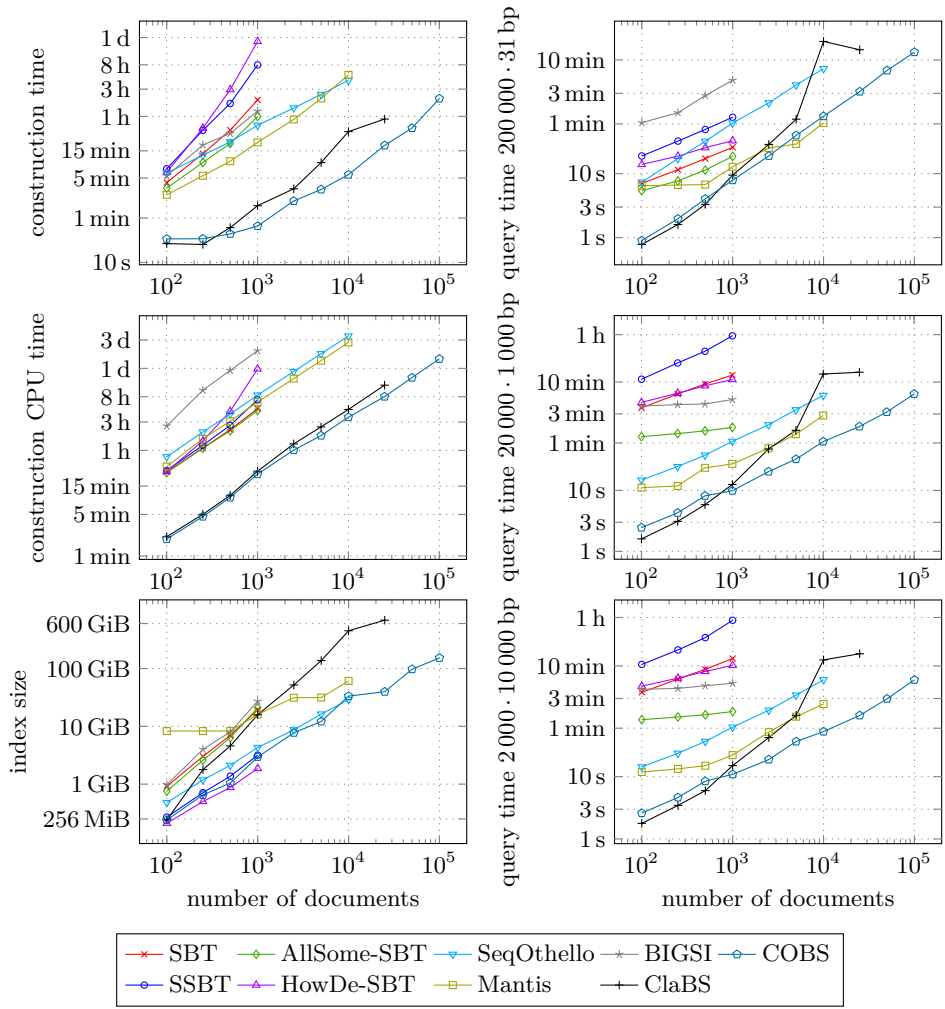
\begin{figure}[t]
  \hspace{11ex}
  \begin{tikzpicture}[trim axis left]
    \begin{loglogaxis}[myPlot,
      ylabel={construction time},
      ytick=\empty,
      extra y ticks={10,60,300,900,3600,10800,28800,86400},
      extra y tick labels={10\,s,1\,min,5\,min,15\,min,1\,h,3\,h,8\,h,1\,d},
      ]

      \addplot coordinates { (100,249.91) (250,803.44) (500,2072.04) (1000,6995.78) };
      \addlegendentry{SBT};
      \addplot coordinates { (100,434.61) (250,2058.41) (500,6066.2) (1000,28657.24) };
      \addlegendentry{SSBT};
      \addplot coordinates { (100,200.06) (250,562.55) (500,1208.49) (1000,3575.87) };
      \addlegendentry{AllSome-SBT};
      \addplot coordinates { (100,379.1) (250,2243.93) (500,10606.58) (1000,73939.13) };
      \addlegendentry{HowDe-SBT};
      \addplot coordinates { (100,366.62) (250,753.86) (500,1289.06) (1000,2520.28) (2500,5053.93) (5000,8730.51) (10000,15379.68) };
      \addlegendentry{SeqOthello};
      \addplot coordinates { (100,152.71) (250,329.34) (500,591.58) (1000,1263.73) (2500,3187.25) (5000,7396.71) (10000,19216.71) };
      \addlegendentry{Mantis};
      \addplot coordinates { (100,327.32) (250,1138.38) (500,1813.89) (1000,4454.52) };
      \addlegendentry{BIGSI};
      \addplot coordinates { (100,21.4) (250,20.69) (500,40.75) (1000,98.57) (2500,194.92) (5000,560.86) (10000,1950.92) (25000,3236.7) };
      \addlegendentry{ClaBS};
      \addplot coordinates { (100,25.87) (250,26.02) (500,31.7) (1000,43.39) (2500,119.26) (5000,189.41) (10000,344.48) (25000,1117.44) (50000,2248.52) (100000,7373.25) };
      \addlegendentry{COBS};

      \legend{}
    \end{loglogaxis}
  \end{tikzpicture}
  \hfill
  \begin{tikzpicture}
    \begin{loglogaxis}[myPlot,
      ylabel style={align=center,overlay},
      ylabel={query time 200\,000\,$\cdot$\,31\,bp},
      ytick=\empty,
      extra y ticks={1,3,10,60,180,600,3600},
      extra y tick labels={1\,s,3\,s,10\,s,1\,min,3\,min,10\,min,1\,h},
      ]

      \addplot coordinates { (100,7.04) (250,11.51) (500,17.2) (1000,25.87) };
      \addlegendentry{SBT};
      \addplot coordinates { (100,18.9) (250,32.29) (500,49.05) (1000,76.27) };
      \addlegendentry{SSBT};
      \addplot coordinates { (100,5.4) (250,7.71) (500,11.37) (1000,18.69) };
      \addlegendentry{AllSome-SBT};
      \addplot coordinates { (100,13.92) (250,18.76) (500,25.47) (1000,32.68) };
      \addlegendentry{HowDe-SBT};
      \addplot coordinates { (100,7.34) (250,17.17) (500,32.02) (1000,61.88) (2500,127.87) (5000,243.34) (10000,437.89) };
      \addlegendentry{SeqOthello};
      \addplot coordinates { (100,6.48) (250,6.63) (500,6.74) (1000,12.79) (2500,25.1) (5000,29.03) (10000,62.13) };
      \addlegendentry{Mantis};
      \addplot coordinates { (100,62.72) (250,89.58) (500,166.3) (1000,288.64) };
      \addlegendentry{BIGSI};
      \addplot coordinates { (100,0.78) (250,1.6) (500,3.31) (1000,9.47) (2500,28.65) (5000,71.09) (10000,1167.87) (25000,865.45) };
      \addlegendentry{ClaBS};
      \addplot coordinates { (100,0.9) (250,1.97) (500,4.03) (1000,7.89) (2500,18.93) (5000,39.61) (10000,79.01) (25000,191.86) (50000,410.6) (100000,793.94) };
      \addlegendentry{COBS};

      \legend{}
    \end{loglogaxis}
  \end{tikzpicture}

  \vspace{-2ex}
  \hspace{11ex}
  \begin{tikzpicture}[trim axis left]
    \begin{loglogaxis}[myPlot,
      ylabel={construction CPU time},
      ytick=\empty,
      extra y ticks={10,60,300,900,3600,10800,28800,86400,259200,691200},
      extra y tick labels={10\,s,1\,min,5\,min,15\,min,1\,h,3\,h,8\,h,1\,d,3\,d,8\,d},
      ]

      \addplot coordinates { (100,1563.15) (250,3929.42) (500,8025.47) (1000,18131.28) };
      \addlegendentry{SBT};
      \addplot coordinates { (100,1647.5) (250,4375.49) (500,9490.83) (1000,25488.92) };
      \addlegendentry{SSBT};
      \addplot coordinates { (100,1509.38) (250,3924.9) (500,7673.95) (1000,16821.28) };
      \addlegendentry{AllSome-SBT};
      \addplot coordinates { (100,1586.51) (250,5200.2) (500,16321.86) (1000,84494.91) };
      \addlegendentry{HowDe-SBT};
      \addplot coordinates { (100,2806.84) (250,7322) (500,14661.62) (1000,30619.52) (2500,76462.61) (5000,152590) (10000,302279) };
      \addlegendentry{SeqOthello};
      \addplot coordinates { (100,1926.9) (250,5631.62) (500,11123.43) (1000,23527.31) (2500,58122.18) (5000,116437) (10000,236293.9) };
      \addlegendentry{Mantis};
      \addplot coordinates { (100,9256.21) (250,36994.43) (500,79729.76) (1000,171113) };
      \addlegendentry{BIGSI};
      \addplot coordinates { (100,126.21) (250,301.22) (500,628.1) (1000,1603.67) (2500,4625.68) (5000,8939.49) (10000,17583.16) (25000,45029.81) };
      \addlegendentry{ClaBS};
      \addplot coordinates { (100,115.37) (250,275.98) (500,579.6) (1000,1430.3) (2500,3632.14) (5000,6381.02) (10000,13032.07) (25000,28894.58) (50000,60207.3) (100000,124530) };
      \addlegendentry{COBS};

      \legend{}
    \end{loglogaxis}
  \end{tikzpicture}
  \hfill
  \begin{tikzpicture}
    \begin{loglogaxis}[myPlot,
      ylabel style={align=center},
      ylabel={query time 20\,000\,$\cdot$\,1\,000\,bp},
      ytick=\empty,
      extra y ticks={1,3,10,60,180,600,3600},
      extra y tick labels={1\,s,3\,s,10\,s,1\,min,3\,min,10\,min,1\,h},
      ]

      \addplot coordinates { (100,225.81) (250,382.31) (500,556.94) (1000,781.08) };
      \addlegendentry{SBT};
      \addplot coordinates { (100,667.64) (250,1233.7) (500,1921.25) (1000,3435.17) };
      \addlegendentry{SSBT};
      \addplot coordinates { (100,76.25) (250,85.6) (500,94.81) (1000,108.06) };
      \addlegendentry{AllSome-SBT};
      \addplot coordinates { (100,278.18) (250,392.82) (500,520.53) (1000,659.29) };
      \addlegendentry{HowDe-SBT};
      \addplot coordinates { (100,14.76) (250,24.61) (500,37.88) (1000,63.81) (2500,119.4) (5000,210.22) (10000,362.06) };
      \addlegendentry{SeqOthello};
      \addplot coordinates { (100,11.1) (250,11.82) (500,23.47) (1000,27.23) (2500,49.29) (5000,83.5) (10000,169.38) };
      \addlegendentry{Mantis};
      \addplot coordinates { (100,241.33) (250,256.07) (500,261.63) (1000,310.26) };
      \addlegendentry{BIGSI};
      \addplot coordinates { (100,1.6) (250,3.1) (500,5.83) (1000,12.51) (2500,47.7) (5000,96.9) (10000,812.49) (25000,868.8) };
      \addlegendentry{ClaBS};
      \addplot coordinates { (100,2.45) (250,4.29) (500,8.15) (1000,9.86) (2500,20.34) (5000,32.55) (10000,63.38) (25000,111.73) (50000,193.11) (100000,382.72) };
      \addlegendentry{COBS};

      \legend{}
    \end{loglogaxis}
  \end{tikzpicture}

  \hspace{11ex}
  \begin{tikzpicture}[trim axis left]
    \begin{loglogaxis}[myPlot,
      xlabel={number of documents},
      ylabel={index size},
      ytick=\empty,
      extra y ticks={0.25,1,10,100,600},
      extra y tick labels={256\,MiB,1\,GiB,10\,GiB,100\,GiB,600\,GiB},
      ]

      \addplot coordinates { (100,0.914909) (250,3.01711) (500,6.9856) (1000,19.379) };
      \addlegendentry{SBT};
      \addplot coordinates { (100,0.269196) (250,0.710678) (500,1.37293) (1000,3.17837) };
      \addlegendentry{SSBT};
      \addplot coordinates { (100,0.755123) (250,2.6) (500,6.39744) (1000,20.8353) };
      \addlegendentry{AllSome-SBT};
      \addplot coordinates { (100,0.210587) (250,0.500565) (500,0.875679) (1000,1.86716) };
      \addlegendentry{HowDe-SBT};
      \addplot coordinates { (100,0.479736) (250,1.1829) (500,2.11526) (1000,4.30739) (2500,8.69841) (5000,16.2987) (10000,28.9281) };
      \addlegendentry{SeqOthello};
      \addplot coordinates { (100,8.28316) (250,8.28625) (500,8.29494) (1000,16.1) (2500,31.2626) (5000,31.4473) (10000,61.0005) };
      \addlegendentry{Mantis};
      \addplot coordinates { (100,0.977234) (250,3.99974) (500,8.0271) (1000,27.1434) };
      \addlegendentry{BIGSI};
      \addplot coordinates { (100,0.240883) (250,1.78561) (500,4.58191) (1000,15.8564) (2500,51.9884) (5000,137.258) (10000,450.558) (25000,686.289) };
      \addlegendentry{ClaBS};
      \addplot coordinates { (100,0.239597) (250,0.651833) (500,1.04406) (1000,2.95192) (2500,7.72922) (5000,12.1383) (10000,33.3397) (25000,39.675) (50000,97.4167) (100000,152.594) };
      \addlegendentry{COBS};

      \legend{}
    \end{loglogaxis}
  \end{tikzpicture}
  \hfill
  \begin{tikzpicture}
    \begin{loglogaxis}[myPlot,
      legend columns=2,
      transpose legend=true,
      legend to name={leg:by-query-time},
      legend image code/.code={
        \draw[mark repeat=2,mark phase=2] plot coordinates {
          (0cm,0cm) (0.25cm,0cm) (0.5cm,0cm)
        };
      },
      xlabel={number of documents},
      ylabel style={align=center,xshift=-2ex,overlay},
      ylabel={query time 2\,000\,$\cdot$\,10\,000\,bp},
      ytick=\empty,
      extra y ticks={1,3,10,60,180,600,3600},
      extra y tick labels={1\,s,3\,s,10\,s,1\,min,3\,min,10\,min,1\,h},
      ]

      \addplot coordinates { (100,226.21) (250,368.79) (500,529.46) (1000,789.78) };
      \addlegendentry{SBT};
      \addplot coordinates { (100,633.65) (250,1082.75) (500,1709.62) (1000,3243.47) };
      \addlegendentry{SSBT};
      \addplot coordinates { (100,82.15) (250,90.79) (500,98.71) (1000,111.12) };
      \addlegendentry{AllSome-SBT};
      \addplot coordinates { (100,284.09) (250,381.89) (500,489.03) (1000,612.51) };
      \addlegendentry{HowDe-SBT};
      \addplot coordinates { (100,14.36) (250,23.73) (500,36.81) (1000,62.32) (2500,116.77) (5000,204.08) (10000,359.72) };
      \addlegendentry{SeqOthello};
      \addplot coordinates { (100,11.88) (250,13.34) (500,14.99) (1000,22.18) (2500,51.95) (5000,91.88) (10000,146.79) };
      \addlegendentry{Mantis};
      \addplot coordinates { (100,253.32) (250,261.56) (500,288.82) (1000,315.7) };
      \addlegendentry{BIGSI};
      \addplot coordinates { (100,1.78) (250,3.45) (500,5.97) (1000,15.05) (2500,42.46) (5000,96.74) (10000,749.92) (25000,941.5) };
      \addlegendentry{ClaBS};
      \addplot coordinates { (100,2.6) (250,4.63) (500,8.45) (1000,10.91) (2500,18.84) (5000,36.68) (10000,52.97) (25000,96.29) (50000,177.91) (100000,358.45) };
      \addlegendentry{COBS};

    \end{loglogaxis}
  \end{tikzpicture}

  \centering\medskip
  \ref{leg:by-query-time}
  \caption{Construction time, index size, query time for 200\,000\,$\cdot$\,31\,bp, 20\,000\,$\cdot$\,1\,000\,bp, and for 2\,000\,$\cdot$\,10\,000\,bp round 2 after disk cache warm-up.}
  \label{fig:scaling-results}
\end{figure}


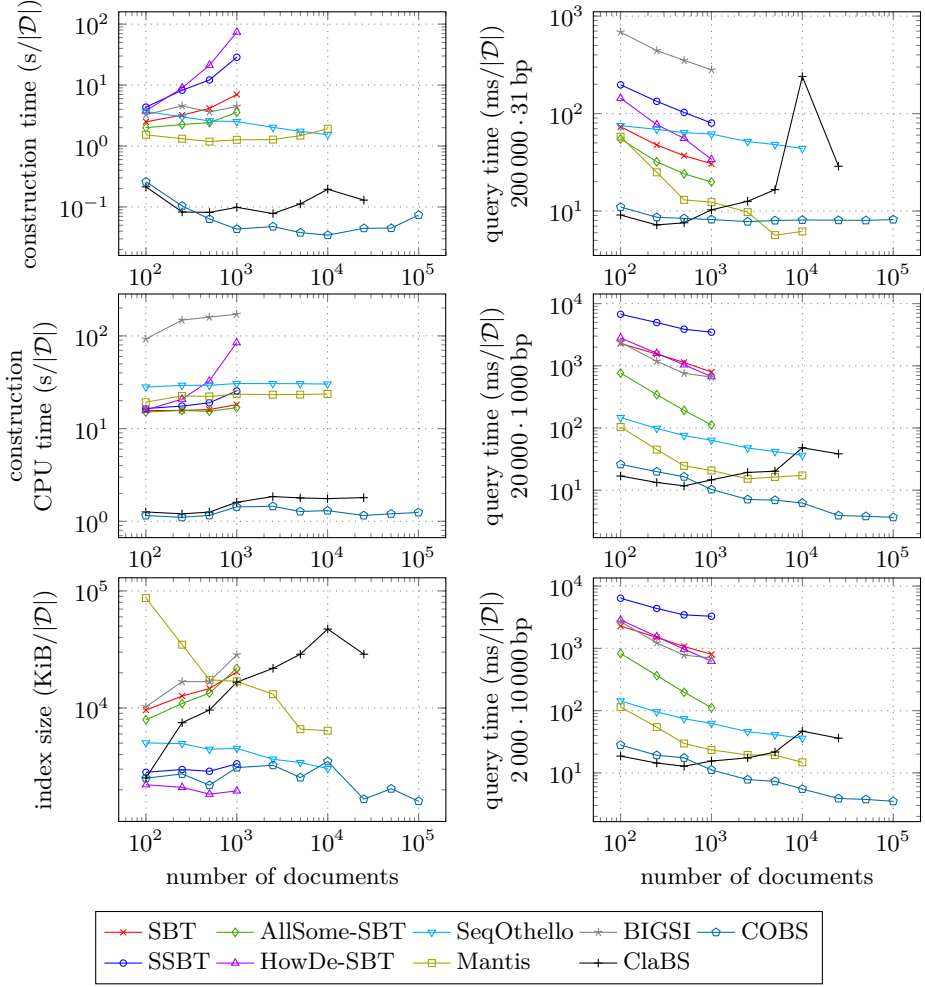
\begin{figure}
  \hspace{11ex}
  \begin{tikzpicture}[trim axis left]
    \begin{loglogaxis}[myPlot,
      ylabel={construction time (s/$|\mathcal{D}|$)},
      ]

      \addplot coordinates { (100,2.4991) (250,3.21376) (500,4.14408) (1000,6.99578) };
      \addlegendentry{SBT};
      \addplot coordinates { (100,4.3461) (250,8.23364) (500,12.1324) (1000,28.65724) };
      \addlegendentry{SSBT};
      \addplot coordinates { (100,2.0006) (250,2.2502) (500,2.41698) (1000,3.57587) };
      \addlegendentry{AllSome-SBT};
      \addplot coordinates { (100,3.791) (250,8.97572) (500,21.21316) (1000,73.93913) };
      \addlegendentry{HowDe-SBT};
      \addplot coordinates { (100,3.6662) (250,3.01544) (500,2.57812) (1000,2.52028) (2500,2.021572) (5000,1.746102) (10000,1.537968) };
      \addlegendentry{SeqOthello};
      \addplot coordinates { (100,1.5271) (250,1.31736) (500,1.18316) (1000,1.26373) (2500,1.2749) (5000,1.479342) (10000,1.921671) };
      \addlegendentry{Mantis};
      \addplot coordinates { (100,3.2732) (250,4.55352) (500,3.62778) (1000,4.45452) };
      \addlegendentry{BIGSI};
      \addplot coordinates { (100,0.214) (250,0.08276) (500,0.0815) (1000,0.09857) (2500,0.077968) (5000,0.112172) (10000,0.195092) (25000,0.129468) };
      \addlegendentry{ClaBS};
      \addplot coordinates { (100,0.2587) (250,0.10408) (500,0.0634) (1000,0.04339) (2500,0.047704) (5000,0.037882) (10000,0.034448) (25000,0.0446976) (50000,0.0449704) (100000,0.0737325) };
      \addlegendentry{COBS};

      \legend{}
    \end{loglogaxis}
  \end{tikzpicture}
  \hfill
  \begin{tikzpicture}
    \begin{loglogaxis}[myPlot,
      ylabel style={align=center},
      ylabel={query time (ms/$|\mathcal{D}|$)\\200\,000\,$\cdot$\,31\,bp},
      ]

      \addplot coordinates { (100,73.3) (250,47.56) (500,36.98) (1000,30.62) };
      \addlegendentry{SBT};
      \addplot coordinates { (100,197.3) (250,133.52) (500,102.7) (1000,80.03) };
      \addlegendentry{SSBT};
      \addplot coordinates { (100,54.9) (250,32.04) (500,24.1) (1000,19.86) };
      \addlegendentry{AllSome-SBT};
      \addplot coordinates { (100,143.6) (250,76.96) (500,55.68) (1000,33.82) };
      \addlegendentry{HowDe-SBT};
      \addplot coordinates { (100,75.2) (250,68.88) (500,63.72) (1000,61.61) (2500,51.66) (5000,48.106) (10000,43.823) };
      \addlegendentry{SeqOthello};
      \addplot coordinates { (100,58) (250,24.96) (500,13.02) (1000,12.35) (2500,9.744) (5000,5.658) (10000,6.18) };
      \addlegendentry{Mantis};
      \addplot coordinates { (100,685.4) (250,442.76) (500,351.42) (1000,281.44) };
      \addlegendentry{BIGSI};
      \addplot coordinates { (100,9.1) (250,7.2) (500,7.58) (1000,10.29) (2500,12.576) (5000,16.588) (10000,240.226) (25000,28.7992) };
      \addlegendentry{ClaBS};
      \addplot coordinates { (100,11) (250,8.64) (500,8.4) (1000,8.19) (2500,7.788) (5000,7.992) (10000,8.097) (25000,8.0552) (50000,8.0202) (100000,8.1802) };
      \addlegendentry{COBS};

      \legend{}
    \end{loglogaxis}
  \end{tikzpicture}

  \vspace{-1ex}
  \hspace{11ex}
  \begin{tikzpicture}[trim axis left]
    \begin{loglogaxis}[myPlot,
      ylabel style={align=center},
      ylabel={construction\\CPU time (s/$|\mathcal{D}|$)},
      ]

      \addplot coordinates { (100,15.6315) (250,15.71768) (500,16.05094) (1000,18.13128) };
      \addlegendentry{SBT};
      \addplot coordinates { (100,16.475) (250,17.50196) (500,18.98166) (1000,25.48892) };
      \addlegendentry{SSBT};
      \addplot coordinates { (100,15.0938) (250,15.6996) (500,15.3479) (1000,16.82128) };
      \addlegendentry{AllSome-SBT};
      \addplot coordinates { (100,15.8651) (250,20.8008) (500,32.64372) (1000,84.49491) };
      \addlegendentry{HowDe-SBT};
      \addplot coordinates { (100,28.0684) (250,29.288) (500,29.32324) (1000,30.61952) (2500,30.585) (5000,30.51801) (10000,30.2279) };
      \addlegendentry{SeqOthello};
      \addplot coordinates { (100,19.269) (250,22.52648) (500,22.24686) (1000,23.52731) (2500,23.2489) (5000,23.28741) (10000,23.62939) };
      \addlegendentry{Mantis};
      \addplot coordinates { (100,92.5621) (250,147.978) (500,159.46) (1000,171.113) };
      \addlegendentry{BIGSI};
      \addplot coordinates { (100,1.2621) (250,1.20488) (500,1.2562) (1000,1.60367) (2500,1.850272) (5000,1.787898) (10000,1.758316) (25000,1.80119) };
      \addlegendentry{ClaBS};
      \addplot coordinates { (100,1.1537) (250,1.10392) (500,1.1592) (1000,1.4303) (2500,1.452856) (5000,1.276204) (10000,1.303207) (25000,1.15578) (50000,1.204146) (100000,1.2453) };
      \addlegendentry{COBS};

      \legend{}
    \end{loglogaxis}
  \end{tikzpicture}
  \hfill
  \begin{tikzpicture}
    \begin{loglogaxis}[myPlot,
      ylabel style={align=center},
      ylabel={query time (ms/$|\mathcal{D}|$)\\20\,000\,$\cdot$\,1\,000\,bp},
      ]

      \addplot coordinates { (100,2285.7) (250,1543.2) (500,1134.14) (1000,794.26) };
      \addlegendentry{SBT};
      \addplot coordinates { (100,6732.1) (250,4962.08) (500,3872.58) (1000,3466.11) };
      \addlegendentry{SSBT};
      \addplot coordinates { (100,765.1) (250,343.08) (500,192.2) (1000,111.83) };
      \addlegendentry{AllSome-SBT};
      \addplot coordinates { (100,2807.7) (250,1589.68) (500,1037.46) (1000,669.85) };
      \addlegendentry{HowDe-SBT};
      \addplot coordinates { (100,147.1) (250,98.6) (500,75.78) (1000,63.48) (2500,47.748) (5000,41.856) (10000,36.38) };
      \addlegendentry{SeqOthello};
      \addplot coordinates { (100,104) (250,44.92) (500,24.64) (1000,20.76) (2500,15.224) (5000,16.312) (10000,17.275) };
      \addlegendentry{Mantis};
      \addplot coordinates { (100,2378.6) (250,1183.08) (500,756.54) (1000,660.06) };
      \addlegendentry{BIGSI};
      \addplot coordinates { (100,16.9) (250,13.32) (500,11.74) (1000,14.68) (2500,19.32) (5000,20.188) (10000,48.121) (25000,38.3948) };
      \addlegendentry{ClaBS};
      \addplot coordinates { (100,26) (250,19.88) (500,16.46) (1000,10.27) (2500,7.1) (5000,6.918) (10000,6.228) (25000,3.9172) (50000,3.8048) (100000,3.6617) };
      \addlegendentry{COBS};

      \legend{}
    \end{loglogaxis}
  \end{tikzpicture}

  \vspace{-1ex}
  \hspace{11ex}
  \begin{tikzpicture}[trim axis left]
    \begin{loglogaxis}[myPlot,
      xlabel={number of documents},
      ylabel={index size (KiB/$|\mathcal{D}|$)},
      ]

      \addplot coordinates { (100,9593) (250,12654) (500,14649) (1000,20320) };
      \addlegendentry{SBT};
      \addplot coordinates { (100,2822) (250,2980) (500,2879) (1000,3332) };
      \addlegendentry{SSBT};
      \addplot coordinates { (100,7918) (250,10905) (500,13416) (1000,21847) };
      \addlegendentry{AllSome-SBT};
      \addplot coordinates { (100,2208) (250,2099) (500,1836) (1000,1957) };
      \addlegendentry{HowDe-SBT};
      \addplot coordinates { (100,5030) (250,4961) (500,4436) (1000,4516) (2500,3648) (5000,3418) (10000,3033) };
      \addlegendentry{SeqOthello};
      \addplot coordinates { (100,86855) (250,34755) (500,17395) (1000,16882) (2500,13112) (5000,6594) (10000,6396) };
      \addlegendentry{Mantis};
      \addplot coordinates { (100,10247) (250,16776) (500,16834) (1000,28461) };
      \addlegendentry{BIGSI};
      \addplot coordinates { (100,2525) (250,7489) (500,9608) (1000,16626) (2500,21805) (5000,28785) (10000,47244) (25000,28785) };
      \addlegendentry{ClaBS};
      \addplot coordinates { (100,2512) (250,2733) (500,2189) (1000,3095) (2500,3241) (5000,2545) (10000,3495) (25000,1664) (50000,2042) (100000,1600) };
      \addlegendentry{COBS};

      \legend{}
    \end{loglogaxis}
  \end{tikzpicture}
  \hfill
  \begin{tikzpicture}
    \begin{loglogaxis}[myPlot,
      legend columns=2,
      transpose legend=true,
      legend to name={leg:by-query-time},
      legend image code/.code={
        \draw[mark repeat=2,mark phase=2] plot coordinates {
          (0cm,0cm) (0.25cm,0cm) (0.5cm,0cm)
        };
      },
      xlabel={number of documents},
      ylabel style={align=center},
      ylabel={query time (ms/$|\mathcal{D}|$)\\2\,000\,$\cdot$\,10\,000\,bp},
      ]

      \addplot coordinates { (100,2280.2) (250,1502.4) (500,1069.22) (1000,802.13) };
      \addlegendentry{SBT};
      \addplot coordinates { (100,6366.1) (250,4364.6) (500,3447.94) (1000,3273.19) };
      \addlegendentry{SSBT};
      \addplot coordinates { (100,827.5) (250,365.56) (500,197.36) (1000,112.03) };
      \addlegendentry{AllSome-SBT};
      \addplot coordinates { (100,2864.4) (250,1544.84) (500,968.82) (1000,621.5) };
      \addlegendentry{HowDe-SBT};
      \addplot coordinates { (100,143.3) (250,95.48) (500,74.1) (1000,62.24) (2500,45.952) (5000,41.092) (10000,35.92) };
      \addlegendentry{SeqOthello};
      \addplot coordinates { (100,113.9) (250,54.28) (500,29.68) (1000,23.38) (2500,19.464) (5000,19.318) (10000,14.776) };
      \addlegendentry{Mantis};
      \addplot coordinates { (100,2688.8) (250,1220.68) (500,779.12) (1000,699.35) };
      \addlegendentry{BIGSI};
      \addplot coordinates { (100,18.6) (250,14.4) (500,12.8) (1000,15.5) (2500,17.424) (5000,21.634) (10000,46.759) (25000,36.1228) };
      \addlegendentry{ClaBS};
      \addplot coordinates { (100,28) (250,19.24) (500,17.42) (1000,11.15) (2500,7.804) (5000,7.332) (10000,5.511) (25000,3.8848) (50000,3.7526) (100000,3.4955) };
      \addlegendentry{COBS};

    \end{loglogaxis}
  \end{tikzpicture}

  \centering\medskip
  \ref{leg:by-query-time}
  \caption{Construction time, index size, query time for 200\,000\,$\cdot$\,31\,bp, 20\,000\,$\cdot$\,1\,000\,bp, and for 2\,000\,$\cdot$\,10\,000\,bp in the first round divided by the number of documents $|\mathcal{D}|$.}
  \label{fig:scaling-results-per-document}
\end{figure}

Considering construction wall-clock time, COBS is clearly the fastest index taking only 43 seconds on 1\,000 documents.
ClaBS is a factor 2.3 slower, Mantis a factor 30 slower, SeqOthello a 59 factor, and AllSome-SBT a factor 83 slower than COBS.
The same is reflected in construction CPU time, with COBS being fastest and taking 1430 seconds.
ClaBS is a factor 1.12 slower, AllSome-SBT a factor 11.8 slower, and Mantis a factor 16.5.

One can also see that we parallelized the Bloom filter construction (the ``bloom'' row) effectively for all indexes, while the build steps are usually only partially parallelized.
COBS has a CPU/wall-clock speedup of 33, while BIGSI has 38, Mantis has 18, and SeqOthello 12.
However, since COBS performs \emph{the least amount} of computation and has among the highest speedups, the combination of these two factors really diminishes wall-clock construction time.
Considering CPU user time, the index requiring most work for construction is BIGSI, probably due to the Python implementation.
It however is parallelized, such that the wall-clock time is on par with the SBTs.

The amount of RAM required by the indexing software also limits their applicability, especially if the complete index itself needs to be constructed in RAM.
BIGSI, HowDe-SBT, and Mantis have the highest main memory usage in the experiment.
For BIGSI and Mantis memory was the limiting scalability factor, while for HowDe-SBT the construction time grew too long.

The index sizes of all packages for the 1\,000 microbial documents was smaller than the input in McCortex format (41\,GiB input size).
The software with the smallest index was the HowDe-SBT with only 1.9\,GiB, followed by COBS with around 3.0\,GiB and SSBT with 3.3\,GiB.

In terms of query performance, the fastest index was COBS with 114 seconds to run all query sets three times, followed by ClaBS with 154 seconds.
Mantis was a factor 2.2 slower, SeqOthello a factor 6.9 slower, and the fastest SBT version, AllSome-SBT, was a factor 8.8 slower.

Using result checkers we verified that all software packages calculated correct results and counted the false positives contained in the returned list of the single $k$-mer query set ($\ell = 31$).
The notable exception was SeqOthello, which produced false positives consistently for each multi-$k$-mer query and started returning false negative (missing) results when run on the 10\,000 dataset. We could not investigate this issue further.
The SBT variants and SeqOthello showed a very low false positive rate less than 0.5\,\%.
Mantis produced zero false positives as expected.
BIGSI and ClaBS are nearly identical in underlying data structure design, and deliver around 2.6\,\% false positives on single $k$-mer queries.
COBS is designed to deliver about the prescribed error rate of 0.3, hence the 22.7\,\% false positives, which enables us to construct a more compact index.
We also calculated the number of false positives in larger multi-$k$-mer query sets, and found all indexes except SeqOthello but including COBS to return \emph{zero} false positives for all queries with $\ell \geq 100$ in the experiment.

\autoref{fig:scaling-results} shows scaling results for all software packages on increasing subsets of the indexed microbial document set.
We skipped running the SBT variants for data sets larger than 10\,000 because their construction time was growing super-linearly.
SeqOthello and Mantis scaled much better in terms of construction time per document.
\autoref{fig:scaling-results-per-document} shows construction time per document.
These plots show that COBS scales well, with an order of magnitude faster construction time per document than Mantis and SeqOthello, both in wall-clock and CPU time.
While ClaBS's index size appears to increase with the number of documents (due to the maximum document size), the size per document of COBS actually decreases because it can better pack documents into blocks.

As expected COBS' query time for single $k$-mers increases linearly with the number of documents in the index, due to the scoring method without pruning.
The query time of all other indexes also increases with the number of documents, but not quite linearly.
The best index in terms of query time increase per document is the AllSome-SBT followed by HowDe-SBT, but only COBS index scales to our full 100\,000 microbial dataset.

\section{Conclusion and Future Work}

With COBS we presented a signature index based on Bloom filters which enables approximate pattern matching on large $q$-gram datasets.
It outperforms all other $q$-gram indexes in both construction and query time for multi-$q$-gram queries due to its simple data structure.

There are many avenues for future work on possible improvements to COBS' ideas.
For example, dynamic operations on the index such as insertion, replacement, and removal of documents are very important for practical applications.
We already provide a merge operation for classic indexes, but not for compact ones.
Our current COBS implementation also already supports querying of multiple index files, such that a frontend may select different datasets or categories.
Another important topic is better support for batch or bulk queries.
And for further scalability it is important to explore distributed index construction and query processing.

Deriving from the simplicity of COBS are research avenues which could explore compression of rows in the Bloom filter matrix using techniques from information retrieval.
And similar to Mantis' use of the CQF one could explore how to adapt other Bloom filter variants to the indexing problem with allowed false positives.


\bibliographystyle{plain}
\bibliography{references}


\end{document}